\documentclass[a4paper,fleqn,usenatbib]{mnras}
\usepackage{natbib,aas_macros}

\usepackage{newtxtext,newtxmath}

\usepackage[T1]{fontenc}
\usepackage{ae,aecompl}

\usepackage{graphicx}	
\usepackage{amsmath}	
\usepackage{amssymb}	

\title[Scattered-light images of fluffy dust discs]{Effect of dust size and structure on scattered-light images of protoplanetary discs}

\author[R. Tazaki et al.]{
Ryo Tazaki,$^{1}$\thanks{E-mail: rtazaki@astr.tohoku.ac.jp}
H. Tanaka,$^{1}$
T. Muto,$^{2}$
A. Kataoka,$^{3}$
and S. Okuzumi$^{4}$
\\
$^{1}$Astronomical Institute, Graduate School of Science,
Tohoku University, 6-3 Aramaki, Aoba-ku, Sendai 980-8578, Japan\\
$^{2}$Division of Liberal Arts, Kogakuin University, 1-24-2 Nishi-Shinjuku, Shinjuku-ku, Tokyo 163-8677, Japan\\
$^{3}$National Astronomical Observatory of Japan, Mitaka, Tokyo 181-8588, Japan\\
$^{4}$Department of Earth and Planetary Sciences, Tokyo Institute of Technology, Meguro, Tokyo, 152-8551, Japan
}

\date{Accepted XXX. Received YYY; in original form ZZZ}

\pubyear{2019}
 
\begin{document}
\label{firstpage}
\pagerange{\pageref{firstpage}--\pageref{lastpage}}
\maketitle

\begin{abstract}
We study scattered light properties of protoplanetary discs at near-infrared wavelengths for various dust size and structure by performing radiative transfer simulations. We show that different dust structures might be probed by measuring disk polarisation fraction as long as the dust radius is larger than the wavelength. When the radius is larger than observing wavelength, disc scattered light will be highly polarised for highly porous dust aggregates, whereas more compact dust structure tends to show low polarisation fraction. Next, roles of monomer radius and fractal dimension for scattered light colours are studied. We find that, outside the Rayleigh regime, as fractal dimension or monomer radius increases, colours of the effective albedo at near-infrared wavelengths vary from blue to red. Our results imply that discs showing grey or slightly blue colours and high polarisation fraction in near-infrared wavelengths might be explained by the presence of large porous aggregates containing sub-microns sized monomers.
\end{abstract}




\section{Introduction}
The first step of planet formation is coagulation of dust particles in protoplanetary discs. Initial growth of dust particles yields fractal dust aggregates, whose porosity is higher than 99\% \citep{Kempf99, Ormel07, Okuzumi12}. 
According to grain growth studies, porosity of an aggregate in discs is a matter of debates; some studies suggest highly porosity ($>85\%$) \citep[e.g.,][]{Wada08, Suyama08}, others predicts lower porosity ($67 - 85\%$) \citep[e.g.,][]{Blum06}.
In our Solar System, both fractal aggregates and compact particles are found, e.g., in comet 67P/Churyumov-Gerasimenko \citep{Fulle15, Fulle16, Mannel16, Bentley16}. 
In a debris disc, the presence of mildly porous particles (porosity of about $60\%$) is suggested by observations \citep[e.g.,][]{Augereau99}. However, porosity of dust aggregates in a protoplanetary disc is still not clear observationally. Since dust porosity is a key to overcome bouncing barrier \citep{Wada11, Kothe13, Brisset17} and radial drift barrier \citep{Okuzumi12, Kataoka13} in dust growth, characterisation of dust porosity from disc observations may shed light on dust evolution in discs.

Since light scattering properties sensitively depend on size, structure, and composition of dust particles \citep[e.g.,][]{Shen08, Shen09, Tazaki16, Tazaki18, Halder18, Ysard18}, disc scattered light may provide information of dust properties. Since disc scattered light has been commonly detected in optical- and near-infrared-wavelengths, in this study, we investigate a connection between dust properties and near-infrared disc scattered light.

Observed disc scattered light colours are thought to reflect wavelength dependence of the albedo of dust particles \citep[e.g.,][]{Mulders13}. In near-infrared wavelengths, most protoplanetary discs show grey colours in total intensity \citep[e.g.,][]{Fukagawa10}, though some discs have reddish colours \citep[e.g.,][]{Mulders13, Long17}. More recently, \citet{Avenhaus18} found that discs around T-Tauri stars have almost grey colours in polarised intensity.

The relation between radii of compact spherical grains and disc colours has so far been studied \citep[e.g.,][]{Mulders13}. Although \citet{Min12} and \citet{Kirchschlager14} studied how dust structure and grain porosity affect disc scattered light, respectively, these studies did not investigate colours. Hence, a role of dust structure on disc colours has not yet been clarified well. 
Furthermore, \citet{Min12} adopted the effective medium theory to compute optical properties of fluffy aggregates; however, once EMT is applied to fluffy aggregates, it significantly overestimates the degree of forward scattering \citep{Tazaki16, Tazaki18}. Hence, more accurate opacity model should be used. \citet{Kirchschlager14} properly treated optical properties of porous grains; however, they only considered grain porosity up to 60\%, whereas fluffy dust aggregates that may form in discs have much higher porosity, e.g., more than $99\%$ \citep[e.g.,][]{Kataoka13}.

An aim of this study is to examine a relation between disc scattered light (total intensity, polarised intensity, and colours) and dust properties (size and structure). As a dust model, we particularly focus on highly porous dust aggregates (porosity higher than 85\%), whose presence has been suggested by grain growth models \citep{Kempf99, Ormel07, Okuzumi12}. Disc scattered light images are obtained by performing 3D Monte Carlo radiative transfer simulations and optical properties of porous aggregates are obtained by using a rigorous numerical method, the T-Matrix Method \citep{Mackowski96}, or a carefully tested approximate method \citep{Tazaki16, Tazaki18}.

This paper is organised as follows. We first explain, in Section \ref{sec:model}, our star and disc models used in this paper. We also summarise dust particle models and their optical properties. Then, we perform radiative transfer simulations of protoplanetary discs at near-infrared wavelengths and discuss images (Section \ref{sec:image}) and disc scattered light colours (Section \ref{sec:colour}). In Section \ref{sec:why}, we study scattering properties of dust aggregates with various structure. Finally, we compare our results with disc observations and close with a conclusion section.

\section{Models} \label{sec:model}

\subsection{Star and disc model}
The central star is assumed to be a T-Tauri star with the effective temperature of $4000$ K and the radius of $2R_{\odot}$. The dust surface density is assumed to be the single power-law distribution, $\Sigma_d=\Sigma_0(R/R_{\rm out})^{-1}$, where $R$ is the distance from the central star, $\Sigma_0$ is the surface density of dust grains at the outer radius $R_{\rm out}$. The disc inner and outer radii are truncated at $10$ au and $100$ au, respectively. 
The total dust disc mass is set as $10^{-4}M_{\odot}$. 
All simulations shown in this paper has the same total dust disk mass.
The vertical distribution of dust grains is assumed to be the Gaussian distribution
with the dust scale height of $H_d=3.3\times10^{-2} (R/1\ {\rm au})^{\beta}$ au. In this paper, we consider a flared disc geometry with $\beta=1.25$ \citep{Kenyon87}. In Section \ref{sec:gene}, we address how the choice of disc geometry affects our conclusion.

We perform radiative transfer simulations using a 3D Monte Carlo radiative transfer code, \textsc{radmc-3d} \citep{Dullemond12}. 
The radial mesh is uniformly spaced in a logarithmic space of [5,105] au with 256 grids. 
The zenith and azimuthal angles are uniformly spaced in a linear space of 
 $[2\pi/9,7\pi/9]$ and $[0,2\pi]$ with 128 and 256 grids, respectively.
The number of photon packages for the scattering Monte Carlo simulation is $10^9$. 
In this paper, we consider three near-infrared wavelengths: $\lambda=1.1\ \mu$m, $1.6\ \mu$m, and $2.2\ \mu$m.

\subsection{Dust models and methods} \label{sec:dustmodel}
\begin{figure*}
\begin{center}
\includegraphics[height=6.0cm,keepaspectratio]{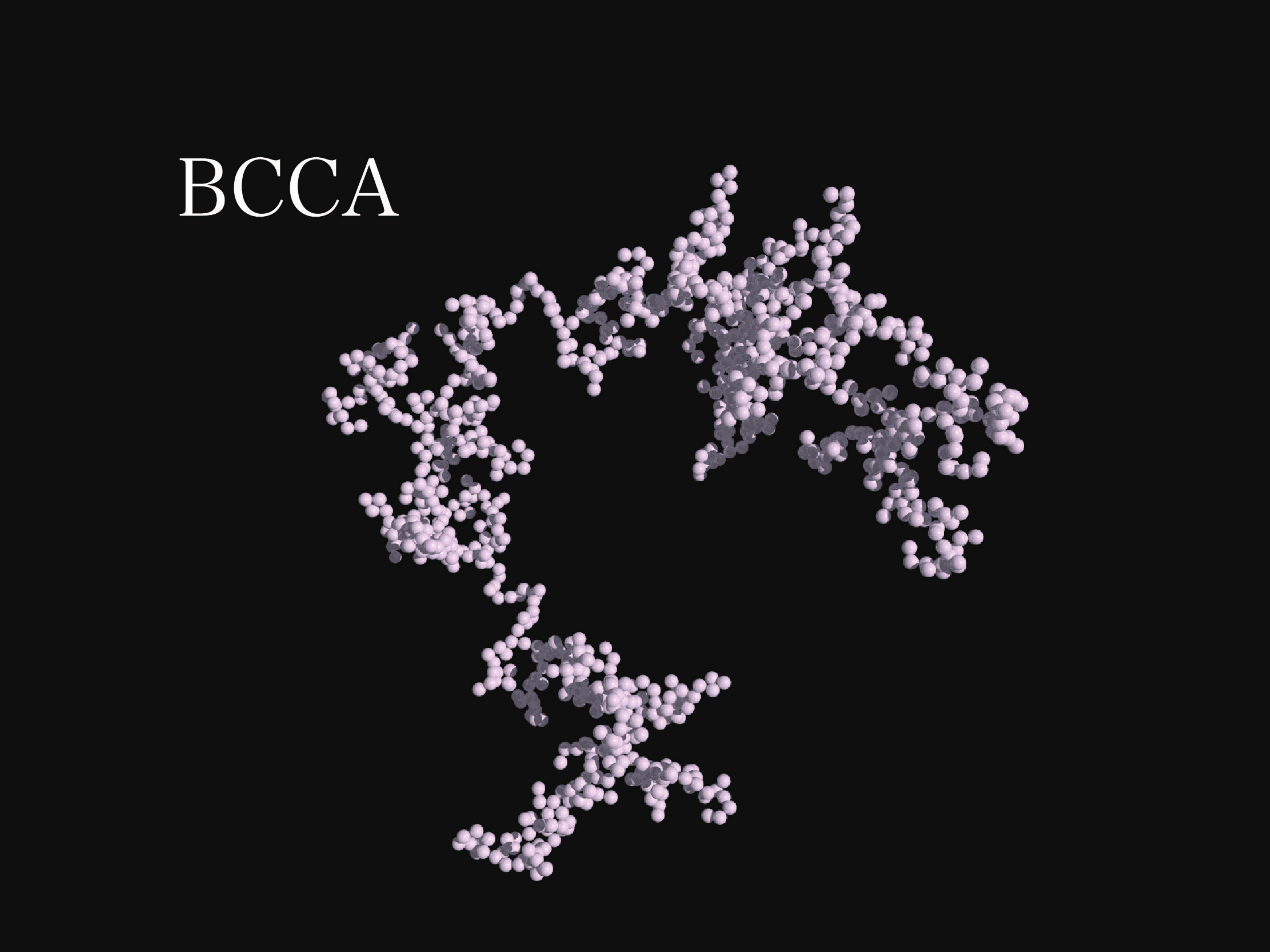}
\includegraphics[height=6.0cm,keepaspectratio]{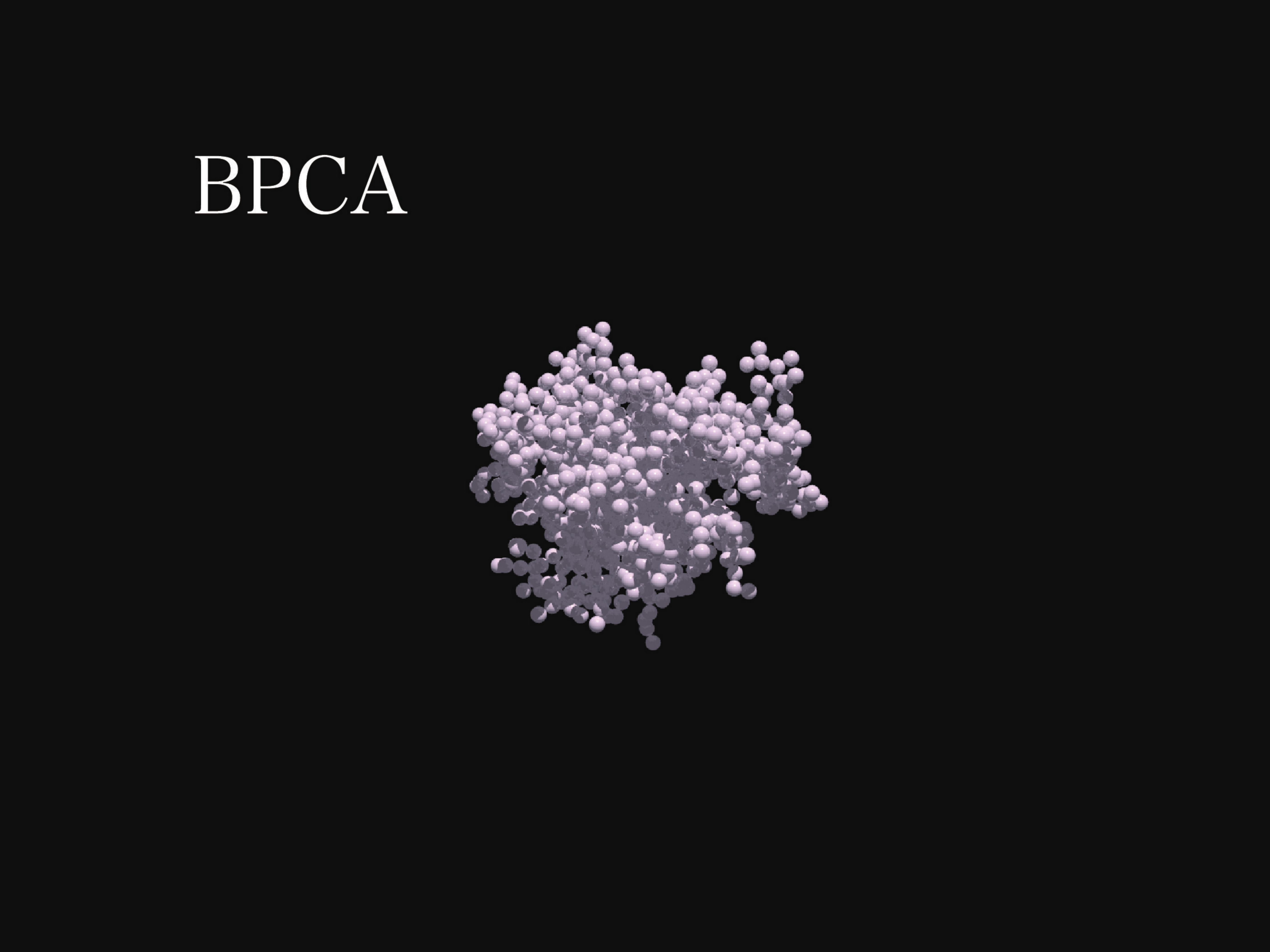}
\caption{Morphology of porous dust aggregates. Left and right panels correspond to the BCCA and BPCA models, respectively. The number of monomers is 1024 and the monomer radius is set as $R_0=0.1\ \mu$m, and hence the characteristic radii of the BCCA and BPCA models are $R_c=4.8\ \mu$m and $1.9\ \mu$m, respectively. Both dust aggregates have the same volume equivalent radius $R_\mathrm{V}\simeq1\ \mu$m. Hence, BCCA and BPCA have porosity of 99\% and 85\%, respectively.}
\label{fig:mor}
\end{center}
\end{figure*}

To study how size and internal structure of dust particles affect observational appearance of discs, we consider three types of dust models. 
In this paper, we basically use the term {\it large} or {\it small} dust radius when the radius is large or small {\it compared to $\lambda/(2\pi)$}, where $\lambda$ is a wavelength of interest.

\begin{itemize}
\item {\it Single monomer model}: 0.1 $\mu$m-sized homogeneous spherical grain (= a monomer particle). This particle is used to form porous dust aggregates described below. Optical properties of the  monomer are obtained by using the Mie theory \citep{Bohren83}. The monomer radius assumed is roughly the same as the constituent particle of cometary dust aggregates \citep[e.g.,][]{Kimura06}. 
Since we assume that a monomer is homogeneous spherical grain, volume filling factor of the monomer is unity.

\item {\it Porous dust aggregate model}: We consider two types of fractal dust aggregates, ballistic cluster cluster agglomerate (BCCA) and ballistic particle cluster agglomerate (BPCA) (see Figure \ref{fig:mor}). Fractal dust aggregates obey a relation $N=k_0(R_g/R_0)^{d_f}$, where $N$ is the number of monomers, $R_g$ is the radius of gyration of the aggregate, $R_0$ is the radius of a monomer, $k_0$ is the fractal prefactor, and $d_f$ is the fractal dimension. Typically, BCCA and BPCA have fractal dimension and fractal prefactor $d_f=1.9$ and $k_0=1.04$ and $d_f=3.0$ and $k_0=0.3$, respectively \citep{Tazaki16}. It is convenient to define the characteristic radius by $R_c=\sqrt{5/3}R_g$ \citep{Mukai92} \footnote{The solid sphere of the radius $a$ has the radius of gyration $R_g=\sqrt{3/5}a$. Thus, characteristic radius of porous aggregate with the radius of gyration $R_g$ is defined in this way.}. In this paper, volume filling factor of the aggregate is defined by $f=NR_0^3/R_c^3$ \citep{Mukai92} \footnote{A way to define filling factor is not unique, and the choice may somewhat affect porosity values, in particular for BCCA. Our definition of filling factor has been widely used \citep[e.g.,][]{Mukai92}, and hence, in this paper, we follow this conventional definition. }. In addition, it is also useful to define volume-equivalent radius $R_\mathrm{V}\equiv R_0N^{1/3}=R_cf^{1/3}$. Porosity of dust aggregates can be defined by $P=1-f$. If the volume filling factor is unity, the characteristic radius is equal to the volume equivalent radius.

Optical properties of the BCCA and BPCA models are obtained by using a rigorous numerical method, T-Matrix Method \footnote{The code is available on \url{ftp://ftp.eng.auburn.edu/pub/dmckwski/scatcodes/}. A newer version of this code is available at \url{http://www.eng.auburn.edu/~dmckwski/scatcodes/}.} \citep[TMM;][]{Mackowski96} with the Quasi-Monte Carlo orientation averaging method \citep{Okada08}\footnote{More detailed information about TMM calculations of porous dust aggregate model is available in \citet{Tazaki16} and \citet{Tazaki18}. }. We also use the modified mean field theory (MMF) developed in \citet{Tazaki18}\footnote{The Rayleigh--Gans--Debye theory studied in \citet{Tazaki16} is a single scattering theory, whereas multiple scattering is often important for relatively large compact aggregates with $d_f>2$. A simple way to implement multiple scattering is to adopt the mean field approximation \citep{Berry86, Botet97}, although the mean field approximation fails to predict the single scattering albedo, once multiple scattering inside the aggregate becomes dominant \citep{Tazaki18}. \citet{Tazaki18} proposed the modified mean field theory in which inaccurate behaviours in the mean field theory are improved by employing geometrical optics approximation \citep[e.g.,][]{Bohren83}.}.

We are mainly interested in porous dust aggregates whose radii are larger than $\lambda/(2\pi)$ because small aggregates simply give rise to Rayleigh scattering. For TMM computation, we adopt $N=1024$. Although $N=1024$ is not a large number, for $R_0=0.1\ \mu$m, the characteristic radii of the BCCA and BPCA models are $R_c=4.8\ \mu$m and $1.9\ \mu$m, respectively; thus, the characteristic radii exceed $\lambda/(2\pi)$ when $\lambda$ corresponds to near-infrared wavelengths.
Porosity of the BCCA and BPCA models is 99\% and 85\%, respectively. For further large aggregates, TMM computation becomes time-consuming, and hence we use MMF instead of TMM. Comparison between MMF and TMM is available in \citet{Tazaki16}, \citet{Tazaki18}, and Appendix \ref{sec:approx}.

\item {\it Compact dust aggregate model}: 
Single-sized compact dust aggregate, whose porosity is low enough so that the distribution of hollow spheres\footnote{In order to avoid strong resonances appeared in scattering properties, we prefer to use the DHS method rather than the Mie theory for compact particles larger than $\lambda/(2\pi)$.} \citep[DHS;][]{Min05, Min16} can be applicable outside the Rayleigh domain. 
We adopt the irregularity parameter $f_{\mathrm{max}}=0.8$ \citep{Min16}.
The volume equivalent radius of a compact dust aggregate is varied from 0.1 $\mu$m to 5.0 $\mu$m. In the DHS method, volume filling factor of an aggregate is not necessary to be specified.
\end{itemize}
Among three dust models, we adopt astronomical silicate for the optical constant \citep{Draine84, Laor93}. In this case, opacities and scattering matrix elements of the BCCA and BPCA models have already presented in \citet{Tazaki16} and \citet{Tazaki18}.
We discuss the effect of dust composition  (Section \ref{sec:gene}), and monomer radius and fractal dimension (Section \ref{sec:why}) in more detail.

\subsection{Optical properties of out dust models} \label{sec:dustoptprop}
\begin{figure*}
\begin{center}
\includegraphics[height=6.0cm,keepaspectratio]{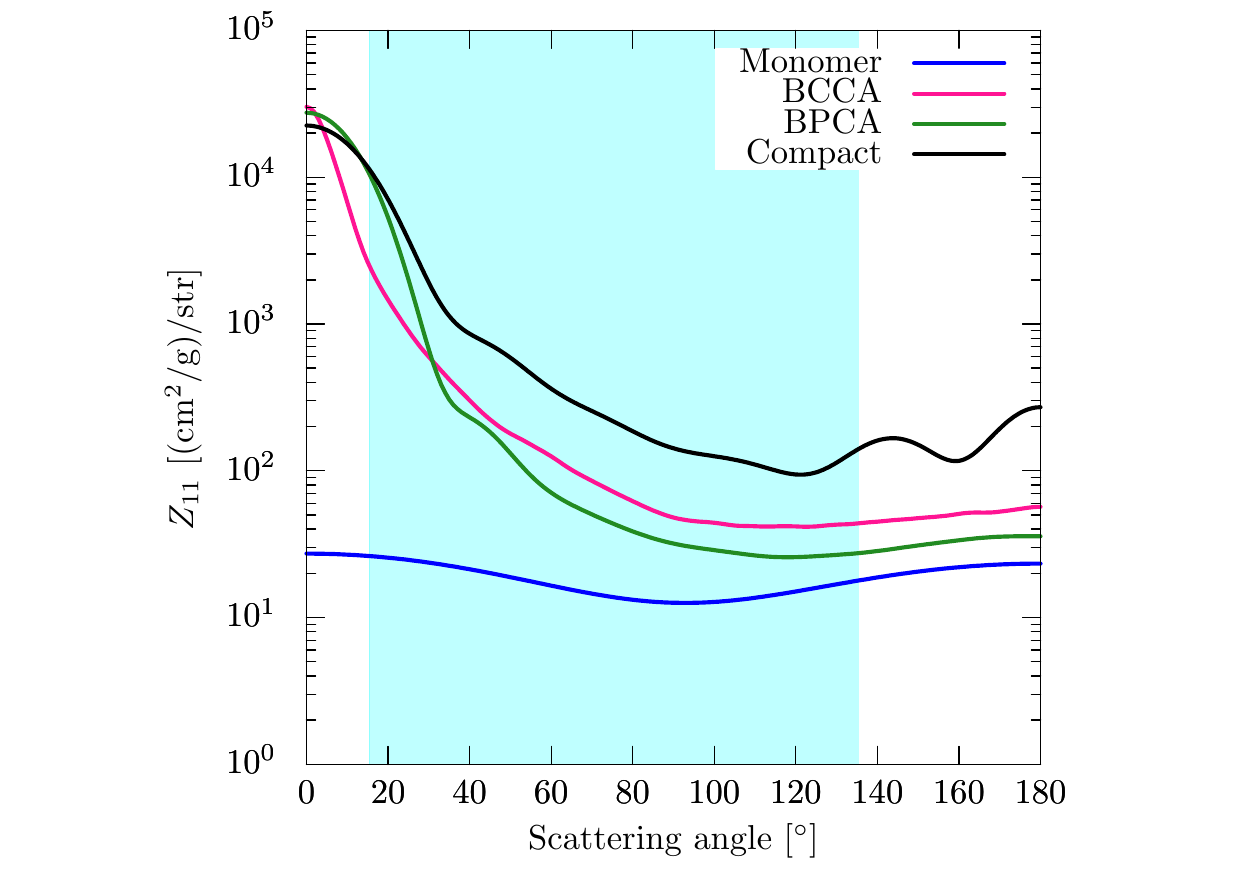}
\includegraphics[height=6.0cm,keepaspectratio]{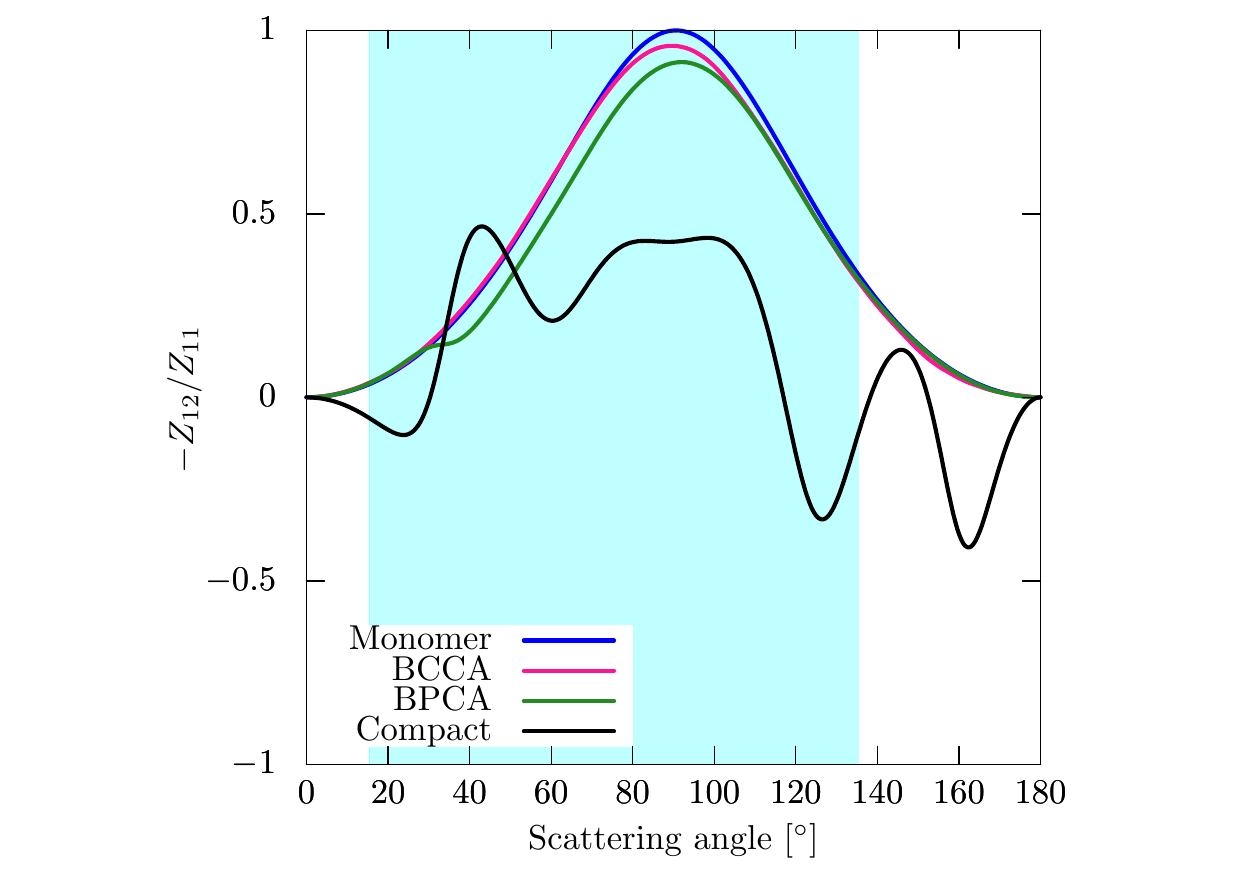}
\caption{
Phase function $Z_{11}$ (left), and the degree of linear polarisation $-Z_{12}/Z_{11}$ (right). Optical properties of the single monomer model, the BCCA and BPCA models, and  the compact dust aggregate model are obtained by using the Mie theory, TMM, and DHS, respectively. A single monomer has the radius of 0.1 $\mu$m, whereas the porous (BCCAs and BPCAs) and compact dust aggregate models have the same volume equivalent radii $1.0\ \mu$m. Wavelength is set as $\lambda=1.6\ \mu$m. Hatched region indicates a range of scattering angle to be observed for a disc with the flaring index $\beta=1.25$ and the inclination angle $i=60^\circ$.}
\label{fig:porousring}
\end{center}
\end{figure*}

We compute optical properties of dust models given in Section \ref{sec:dustmodel}.
Figure \ref{fig:porousring} shows dust optical properties at wavelength $\lambda=1.6\ \mu$m for the single monomer model, the porous dust aggregate model (the BCCA and BPCA models with $N=1024$ and $R_0=0.1\ \mu$m), and the compact dust aggregate model ($R_\mathrm{V}=1\ \mu$m). Since BCCA and BPCA has $N=1024$ and $R_0=0.1\ \mu$m, their volume equivalent radii are $R_\mathrm{V}=R_0N^{1/3}\simeq 1.0\ \mu$m. Therefore, BCCA, BPCA and compact dust aggregates have the same volume equivalent radii.

In this paper, scattering matrix elements are defined by $Z_{ij}=S_{ij}/(k^2m_\mathrm{dust})$, where $S_{ij}$ is those defined by \citet{Bohren83}, $k$ is the wavenumber, and $m_\mathrm{dust}$ is the mass of dust particle. 
Since $Z_{11}$ is a differential cross section per unit mass, it is related to the scattering opacity $\kappa_\mathrm{sca}$ via
\begin{equation}
\kappa_\mathrm{sca}=\oint Z_{11}d\Omega,
\end{equation}
where $d\Omega$ is the solid angle. 
For the sake of simplicity, in this paper, we refer $Z_{11}$ as {\it phase function}.
As shown in Figure \ref{fig:porousring}, scattering is anisotropic for BCCA, BPCA, and compact dust aggregate models. This is because these aggregates have the radii larger than $\lambda/(2\pi)$. On the other hand, the monomer model shows isotropic scattering due to Rayleigh scattering.

The degree of linear polarisation is given by $-Z_{12}/Z_{11}$.
For small particles, such as a monomer particle, the degree of polarisation is 100\% at 90 degree scattering angle due to Rayleigh scattering.
On the other hand, for a particle larger than $\lambda/(2\pi)$, the degree of polarisation depends on the structure of dust aggregates. Porous dust aggregates tend to show high degree of polarisation with a bell-shaped profile (Figure \ref{fig:porousring}). This is because multiple scattering inside the aggregate is suppressed due to its highly porous structure, and then, polarisation properties are determined by those of the monomer \citep[see Eq. (9) in ][]{Tazaki16}. The degree of polarisation of compact dust aggregates is lower than porous dust aggregates due to occurrence of multiple scattering inside the particle. Hence, the degree of polarisation provides valuable information about structure of dust aggregates when the particle radius is larger than wavelength. 

\section{Results of radiative transfer simulations: Images} \label{sec:image}
\begin{figure*}
\begin{center}
\includegraphics[height=13.0cm,keepaspectratio]{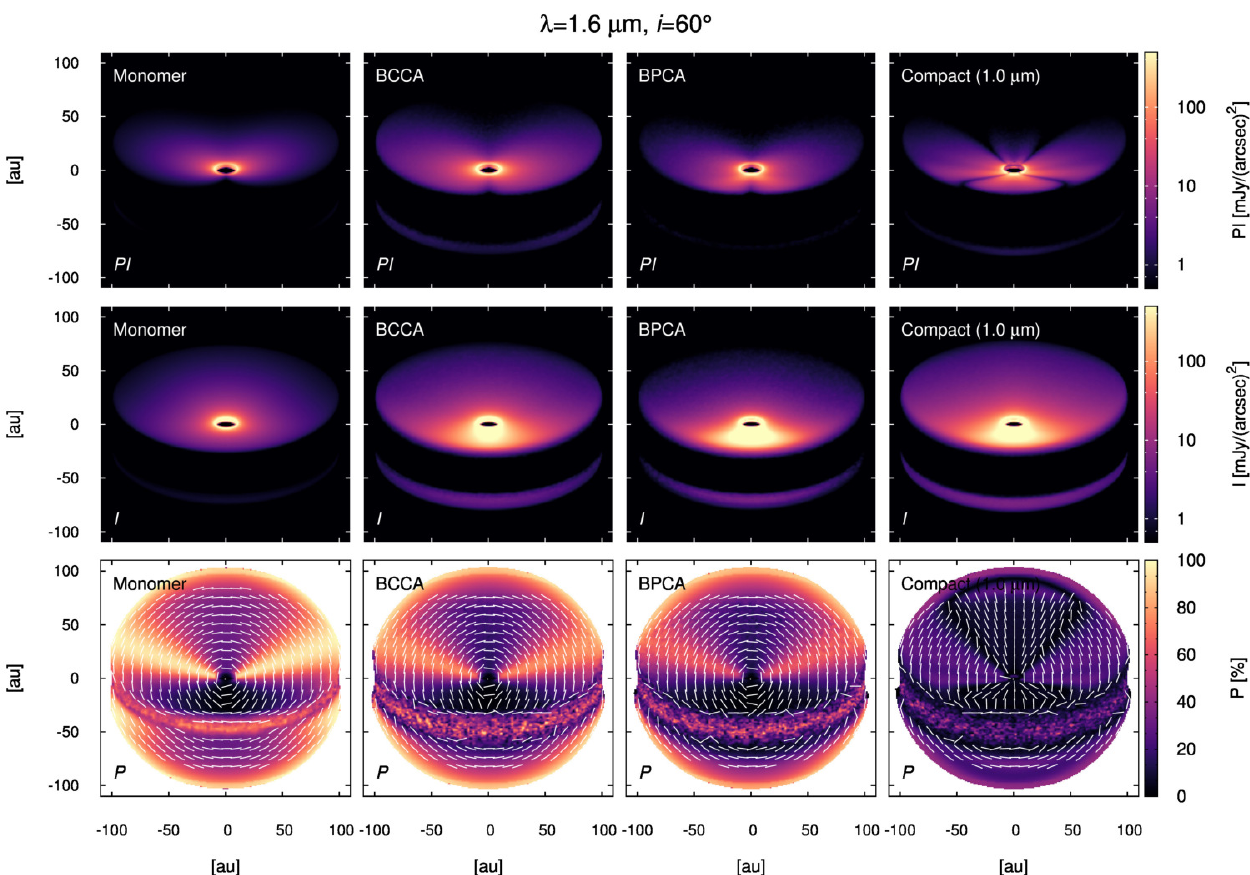}
\caption{Polarised intensity (top), total intensity (middle), and polarisation fraction (bottom). From left to right, each panel indicates the results for the single monomer model (0.1 $\mu$m-sized spherical particles), the BCCA and BPCA models ($N=1024$, $R_0=0.1\ \mu$m, astronomical silicate), and the compact dust aggregate model with the radius of 1.0 $\mu$m. The wavelength is $\lambda=1.6\ \mu$m and the inclination angle is $i=60^\circ$. Optical properties of BCCAs and BPCAs are calculated by a rigorous method, TMM. No star is included in images (perfect coronagraph). For the porous dust aggregate models, the disc shows brightness intensity asymmetry (in total intensity) due to forward scattering as well as high polarisation fraction.}
\label{fig:rtincl60porous}
\end{center}
\end{figure*}

\begin{figure}
\begin{center}
\includegraphics[height=6.0cm,keepaspectratio]{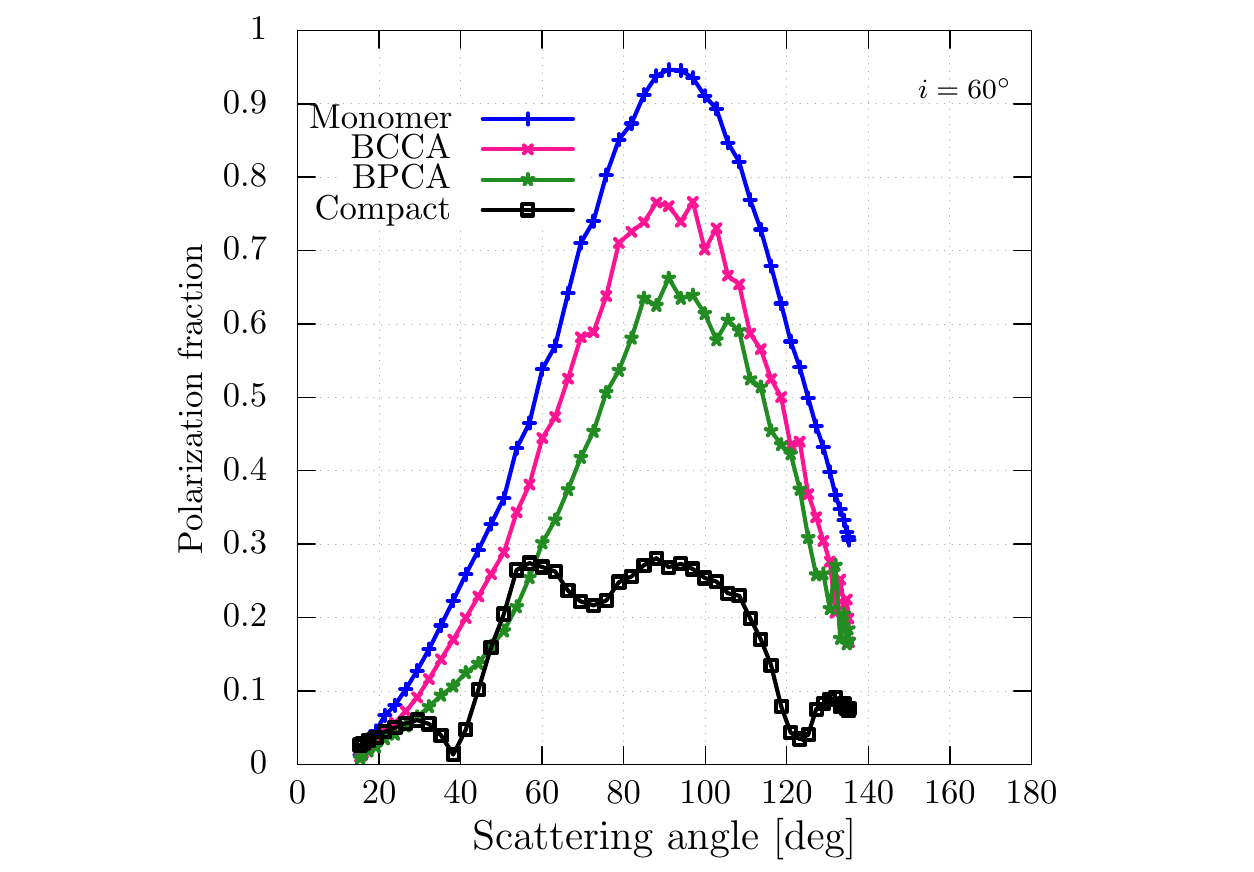}
\caption{Polarisation fraction as a function of scattering angles at $R=50$ au measured from simulated images in Figure \ref{fig:rtincl60porous}. Because of multiple scattering at the disc surface, polarisation fraction is smaller than the degree of polarisation given in Figure \ref{fig:porousring} (right).}
\label{fig:polfrac}
\end{center}
\end{figure}

We perform radiative transfer simulations using a model described in Section \ref{sec:model}.
Optical properties used in the simulations are the ones summarised in Section \ref{sec:dustoptprop}.

Total intensity images (middle panels of Figure \ref{fig:rtincl60porous}) can be used as a diagnostics of particle radius.
Both the porous and compact aggregate models show the brightness asymmetry, that is, the near side of the disc is brighter than the backward side. This is because the dust particle radii assumed are larger than the wavelength (radius $\gtrsim \lambda/2\pi$), and hence strong forward scattering occurs. Meanwhile, for the single monomer model, the brightness asymmetry is weak because the monomer is a Rayleigh scatterer, that is, isotropic scattering occurs. 

Although differences between porous aggregates (BCCA and BPCA) and compact aggregates are less obvious in total intensity images, they clearly differ in images of polarisation fraction.
Figure \ref{fig:polfrac} shows polarisation fraction measured at $R=50$ au from Figure \ref{fig:rtincl60porous}, where scattering angles are estimated by using a scattered-light mapping method \citep{Stolker16b}. 
As shown in Figure \ref{fig:polfrac}, the polarisation fraction of the BCCA and BPCA models is as high as 65\% - 75\%. Polarisation fraction obtained by simulations is somewhat smaller than the degree of polarisation (Figure \ref{fig:porousring}). This is mainly due to occurrence of multiple scattering at the disc surface because the disk is optically thick. When we compare Figures \ref{fig:porousring} and \ref{fig:polfrac}, it is found that the degree of polarisation is similar between the monomer, the BCCA and BPCA models, whereas highest polarisation fraction observed in the discs for these particles can largely differ.

Polarisation fraction of the compact dust aggregate model can be much smaller than those of porous dust aggregate models. This is because multiple scattering easily occurs for more compact structure of dust aggregates once the aggregate radius becomes larger than $\lambda/2\pi$ \footnote{When the particle has refractive index close to unity, polarisation fraction remains high even if the size parameter exceeds unity \citep[see e.g., chapter 6 of][]{Bohren83}. This is due to multiple scattering inside the sphere becomes sub-dominant for such a transparent material.}. As a result, compact dust aggregates show lower polarisation fraction compared to the porous dust aggregates. 

Table \ref{table:pfrac} summarises integrated polarisation fraction of discs for various dust models. For all dust models, more inclined discs show lower polarisation fraction because wider scattering angles can be observed. The BCCA model predicts that integrated polarisation fraction is as high as 63\% for a face-on disc, whereas it is only 14\% for an very inclined disc ($i=75^{\circ}$). On the other hand, more compact dust structure, like BPCA and compact dust models, results in smaller integrated polarisation fraction than those of BCCA. As a result, it is found that disc polarisation fraction is an important quantity to distinguish different dust structure, where relatively high polarisation fraction of BCCA and monomer could be distinguished by the presence of forward scattered light in total intensity image.

\begin{table}
  \caption{Disc Integrated Polarisation Fraction at $\lambda=1.6\ \mu$m for the our Dust Models}
  \label{table:pfrac}
  \centering
  \begin{tabular}{ccccc}
    \hline
    Inclination angle & monomer & BCCA & BPCA & Compact  \\
    \hline \hline
    $i=0^\circ$ & 84\% & 63\%  & 46\% & 22\% \\
    $i=15^\circ$ & 80\% & 57\% & 42\% & 19\%  \\
    $i=30^\circ$ & 68\% & 46\% & 31\% & 21\%  \\
    $i=45^\circ$ & 53\% & 32\% & 19\% & 13\%  \\
    $i=60^\circ$ & 42\% & 18\% & 8\% & 8\%  \\
    $i=75^\circ$ & 37\% & 14\% & 8\% & 6\%  \\
   \hline
 \end{tabular}
\end{table}

The tendency that porous aggregates show high polarisation fraction is compatible with the previous works by \citet{Min12}, who used EMT, and by \citet{Kirchschlager14}, who adopted less porous particle models. 
However, as shown in Appendix \ref{sec:approx}, EMT largely under/over-estimate the scattered-light intensity and the polarisation fraction for large fluffy dust aggregates. This is due to the fact that phase function obtained by EMT shows extremely strong forward scattering, which suppresses both large angle scattering (reducing total intensity) and multiple scattering at disc surface (increasing polarisation fraction). 

Polarised intensity does not show clear brightness asymmetry because polarisation fraction becomes small along the minor axis. The compact dust aggregate model shows some slits in the image of polarised intensity. 
In our compact dust aggregate model, oscillatory behaviour appears in the angular profile of the degree of polarisation due to resonances arising from smooth spherical surfaces (see Figure \ref{fig:porousring}). However, realistic compact dust aggregates are thought to have surface roughness, and hence these resonances would be smeared out \citep[e.g.,][]{Min16}. Therefore, these slits in the image thought to be unrealistic. The reason why a monomer particle, which also has smooth surface, does not show the oscillatory behaviour is simply because Rayleigh scattering occurs, that is, scattering is coherent. It is also worth mentioning that even if the oscillatory behaviour is suppressed, polarisation flip at disk backward side may remain for some dust properties because of the effect called negative polarisation branch \citep[e.g.,][]{Kirchschlager14}. This effect, for example, has been observed for cometary dust particles \citep[e.g.,][]{Kolokolova07}. 

\begin{figure*}
\begin{center}
\includegraphics[height=6.0cm,keepaspectratio]{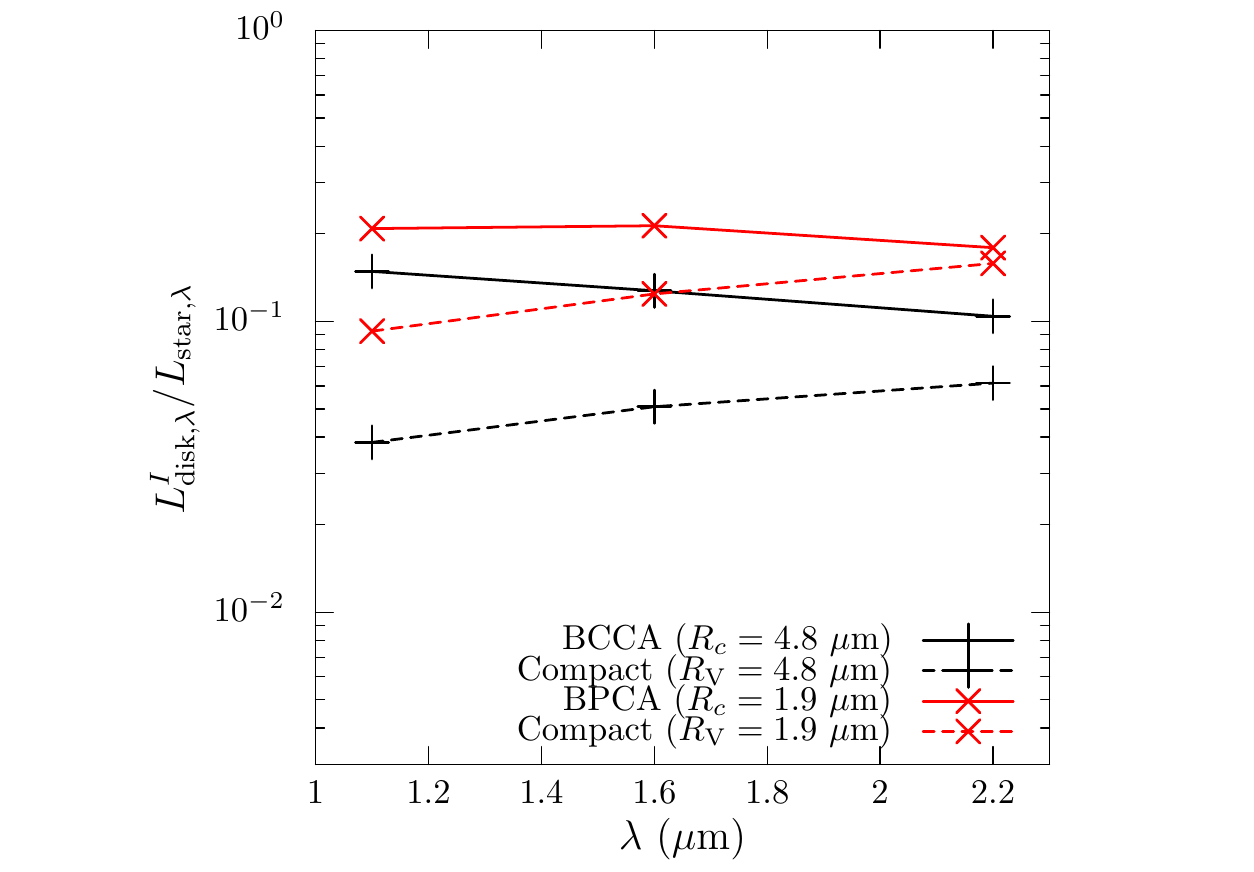}
\includegraphics[height=6.0cm,keepaspectratio]{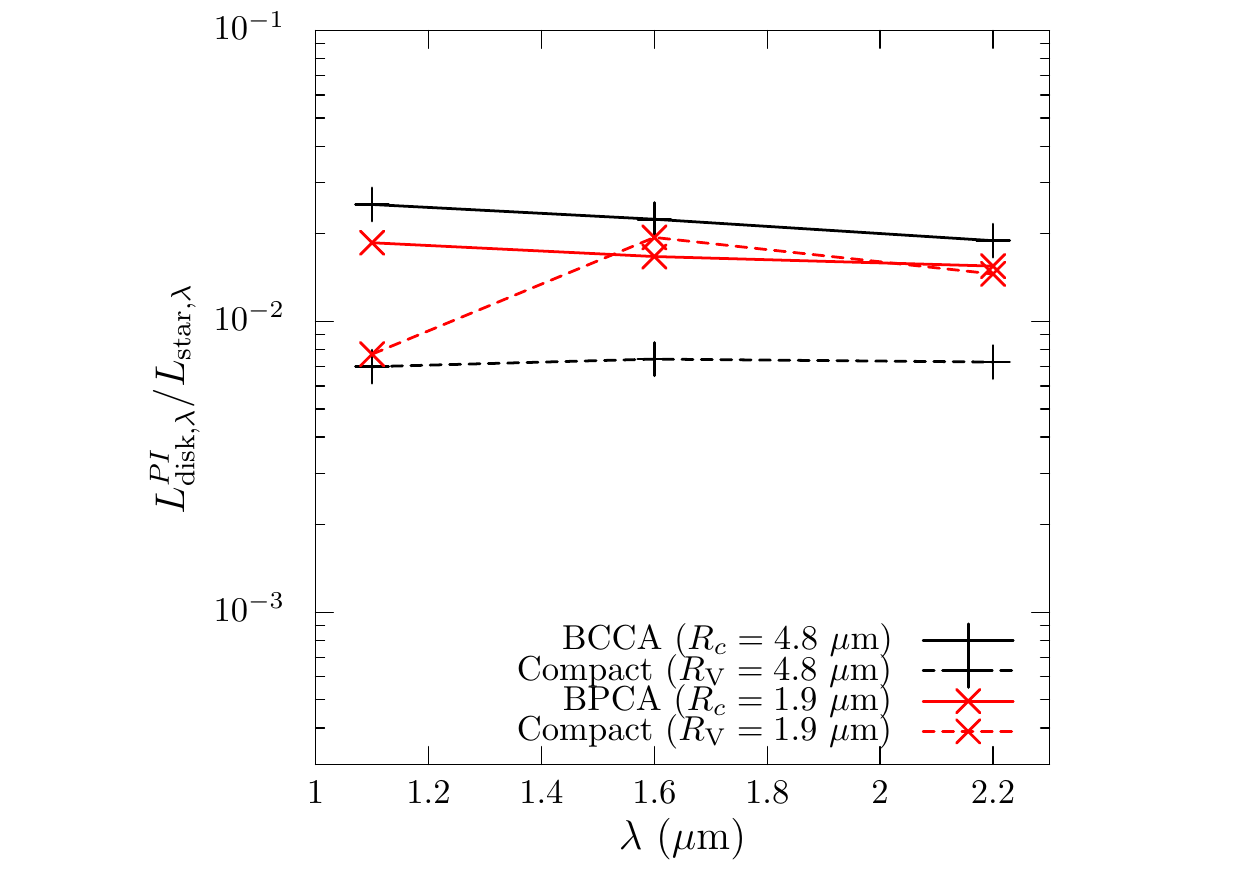}
\caption{Fractional luminosity in total intensity (left) and polarised intensity (right) at the inclination angle $i=60^\circ$. Black and red solid lines represents the results for BCCA and BPCA, respectively. Black and red dashed lines indicate the results for compact dust aggregates with $4.8\ \mu$m and $1.9\ \mu$m radii, respectively.}
\label{fig:colourRc}
\end{center}
\end{figure*}

\section{Results of radiative transfer simulations: Scattered-light colours} \label{sec:colour}
We define the scattered light colour by
\begin{equation}
\eta=\frac{\log(L_{\mathrm{disc},\lambda_2}/L_{\mathrm{star},\lambda_2})-\log(L_{\mathrm{disc},\lambda_1}/L_{\mathrm{star},\lambda_1})}{\log(\lambda_2/\lambda_1)},
\end{equation}
where $L_{\mathrm{disc},\lambda}$ and $L_{\mathrm{star},\lambda}$ are total luminosity of disc and central star, respectively, for either total intensity or polarised intensity. We adopt $\lambda_1=1.1\ \mu$m and $\lambda_2=2.2\ \mu$m. In this paper, we classify scattered-light colours into blue, grey, and red when $\eta<-0.5$, $-0.5\le\eta\le0.5$, and $\eta>0.5$, respectively. 

\subsection{Comapct aggregates vs. porous dust aggregates} \label{sec:compporous}

\begin{table}
  \caption{Disc Scattered Light Colour for Different Aggregate Models}
  \label{table:colour_porous}
  \centering
  \begin{tabular}{lcc}
    \hline
    Dust model  & $\eta_\mathrm{I}$ & $\eta_\mathrm{PI}$ \\
    \hline \hline
    Monomer  & $-2.4$ & $-2.4$ \\
    BCCA ($r_V=1.0\ \mu$m, $R_c=4.8\ \mu$m)  & $-0.51$ & $-0.41$ \\
    BPCA ($r_V=1.0\ \mu$m, $R_c=1.9\ \mu$m)  & $-0.21$ & $-0.27$ \\
    Compact ($R_\mathrm{V}=1.0\ \mu$m) & $0.030$ & $0.099$ \\
    Compact ($R_\mathrm{V}=1.9\ \mu$m) & $0.77$ & $0.91$ \\
    Compact ($R_\mathrm{V}=4.8\ \mu$m) & $0.68$ & $0.053$ \\
    \hline
 \end{tabular}
\end{table}

Figure \ref{fig:colourRc} and Table \ref{table:colour_porous} show scattered-light colour for both in total intensity ($\eta_\mathrm{I}$) and in polarised intensity ($\eta_\mathrm{PI}$)  at inclination angle $i=60^\circ$. 
Figure \ref{fig:colourRc} also compares total intensity colours of BCCA ($R_c=4.8\ \mu$m) to those of the compact aggregate with the same radius. We find that total intensity colours of BCCA ($R_c=4.8\ \mu$m) are grey or slightly blue for both total and polarised intensity. Meanwhile, a compact aggregate with $R_\mathrm{V}=4.8\ \mu$m shows reddish colour \citep[][]{Mulders13}. BPCA ($R_c=1.9\ \mu$m) also show grey colours in total intensity, whereas compact aggregates with 1.9 $\mu$m show reddish colours. As a result, porous aggregates large compared to wavelength tends to show grey or slightly blue colours in total intensity, whereas large compact aggregates give rise to reddish colours. Therefore, even if porous and compact dust aggregates have the same radii, their scattered light colour can differ.
It is worth mentioning that for compact dust aggregates, grey colours can appear when the radius is comparable to the wavelength (see Figure \ref{fig:DHScolour} in Appendix \ref{sec:rtcomp} for more detail).

\subsection{Scattered light colours of millimetre-sized BCCA} \label{sec:exlarge}
Due to strong aerodynamic coupling, highly porous dust aggregates at disc surface may be much larger than micron-size, whereas large compact dust aggregates are likely to settle down to the midplane. 
In Section \ref{sec:compporous}, we concluded that a few micron-sized BCCA are shown to yield grey or slightly blue rather than reddish colours. However, one may doubt that the presence of further larger fluffy aggregates than those considered in Section \ref{sec:compporous} may makes the disc reddish. In this section, we show that colours remain almost the same even if the radius of aggregates is increased to millimetre-size.

\subsubsection{Phase function} \label{sec:exlargeopt}
Since we are interested in aerodynamically coupled aggregates, we consider BCCA, whose mass-to-area ratio is the same as the individual monomer particle.
In particular, we study scattering properties of BCCA with radius from $R_c=1\ \mu$m to $1$ mm. Millimetre-sized BCCA with $R_0=0.1\ \mu$m contains $\sim10^7$ monomers, and hence, TMM computation is time consuming. Thus, we adopt MMF instead of TMM.
For MMF, angular dependence of phase matrix elements are reliably computed when the single scattering assumption is validated \citep{Tazaki18}:
\begin{eqnarray}
\Delta\phi&<&1,\label{eq:cond1}\\
|m-1|&<&2, \label{eq:cond2}
\end{eqnarray}
where $\Delta\phi$ is the (maximum) phase shift of dust aggregates and $m$ is the complex refractive index. For astronomical silicate, the first condition is satisfied for $\lambda\ge0.85\ \mu$m for BCCA with $R_0=0.1\ \mu$m. It should be emphasised that in the case of BCCA, $\Delta\phi$ does not depend on the aggregate radius. In optical and near-infrared wavelengths, the second condition is also satisfied (see Figure 3 in \citet{Tazaki16}). 

\begin{figure}
\begin{center}
\includegraphics[height=6.0cm,keepaspectratio]{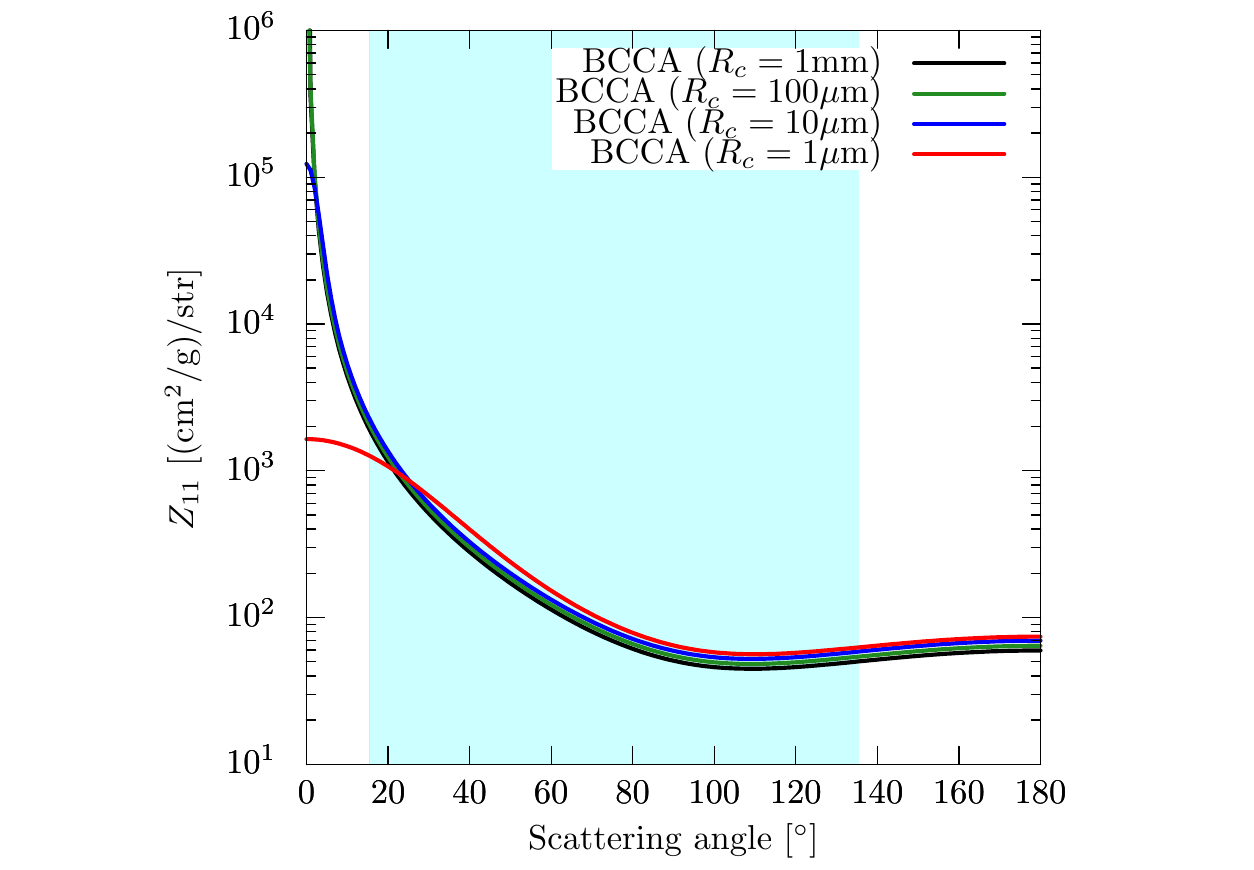}
\caption{Phase function $Z_{11}$ of extremely large BCCA obtained by MMF. The wavelength is set as $\lambda=1.6\ \mu$m. Hatched region indicates a range of scattering angle to be observed for a disc with the flaring index $\beta=1.25$ and the inclination angle $i=60^\circ$.}
\label{fig:Z11large}
\end{center}
\end{figure}

\begin{figure}
\begin{center}
\includegraphics[height=5.0cm,keepaspectratio]{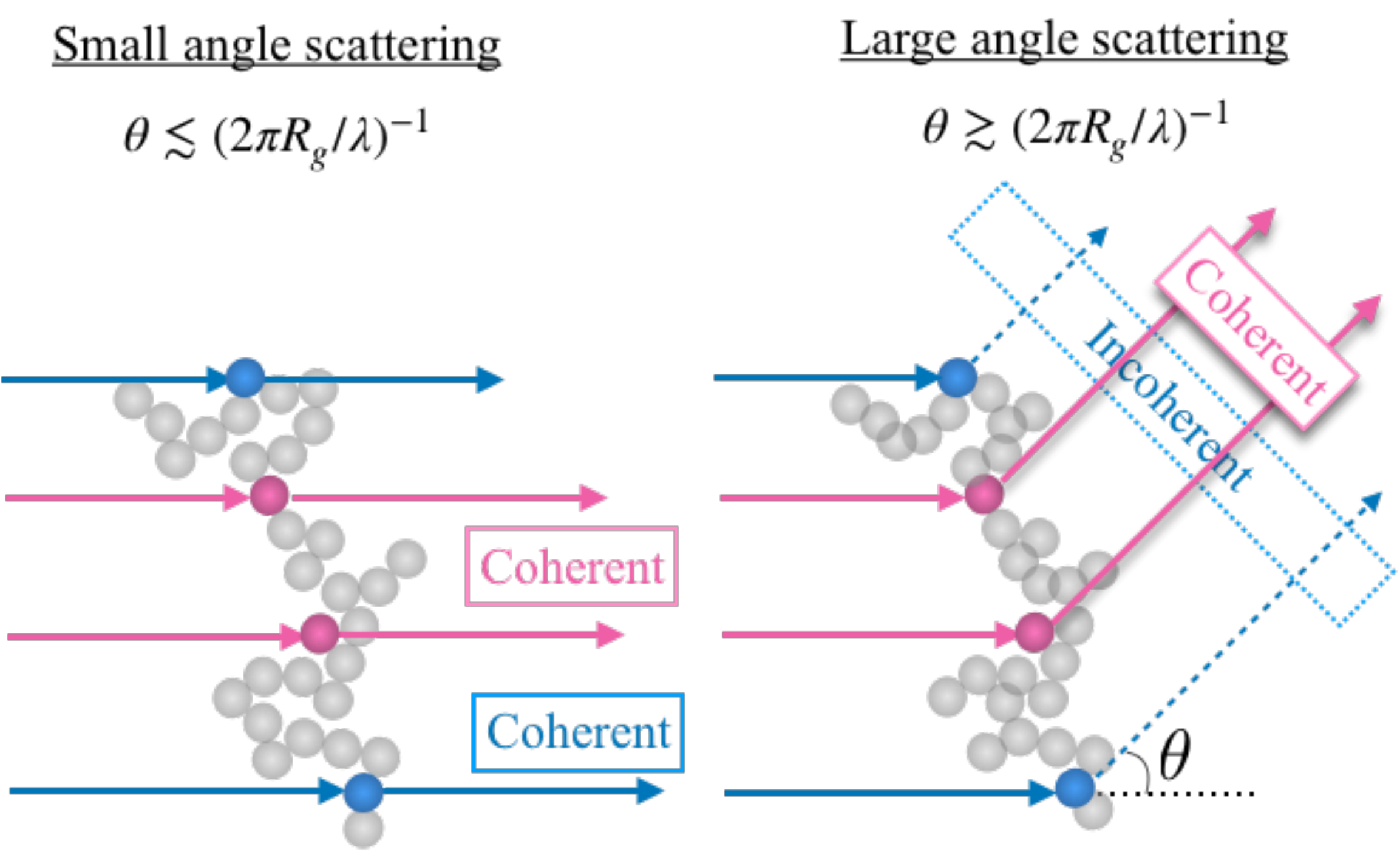}
\caption{Schematic illustration to explain saturation of $Z_{11}$ for large angle scattering. At large scattering angles, scattered light is dominated by coherent light scattered from small scale structure of the aggregate, whereas large scale structure, e.g., aggregate radius, only produces incoherent light due to relatively large optical path of difference. Hence, the aggregate radius is insensitive to $Z_{11}$ at intermediate scattering angles.}
\label{fig:saturation}
\end{center}
\end{figure}

Figure \ref{fig:Z11large} shows phase function $Z_{11}$ of BCCA with radius from $R_c=1\ \mu$m to $1$ mm obtained by MMF. As the aggregate radius increases, forward scattering becomes strong; however, intermediate- and backward-scattered intensity saturates. This saturation is a natural consequence of single scattering by dust aggregates \citep[][]{Berry86, Tazaki16}. Mechanism of the saturation is illustrated in Figure \ref{fig:saturation}.
For large angle scattering ($\theta \gtrsim (2\pi R_g/\lambda)^{-1}$), scattered intensity is dominated by coherent scattered waves from a pair of monomers separated by the distance smaller than the aggregate radius. On the other hand, a pair of monomers with separation as large as aggregate radius only gives rise to incoherent contribution to the scattered intensity because their optical path of difference is large. Thus, scattered intensity at large scattering angles is governed by its small scale structure, e.g., fractal dimension and the monomer's properties, rather than the aggregate radius. As a result, phase function $Z_{11}$ shown in Figure \ref{fig:Z11large} is insensitive to the characteristic radius at intermediate- and back-scattering angles.
For the case of small angle scattering ($\theta \lesssim (2\pi R_g/\lambda)^{-1}$), large scale structure of aggregates can produce coherent scattered light, and therefore, $Z_{11}$ depends on the aggregate radius as can be seen in Figure \ref{fig:Z11large}.

\subsubsection{Disc scattered-light colours}
By using optical properties of BCCA computed in Section \ref{sec:exlargeopt}, we perform radiative transfer simulations of discs. As shown in Figure \ref{fig:Z11large}, millimetre-sized BCCAs show very sharp and intense forward scattering. Hence, a large number of grids will be required in the radiative transfer calculation\footnote{See also Section 6.5.6 in Manual for \textsc{radmc-3d} Version 0.39, which is available on \url{http://www.ita.uni-heidelberg.de/~dullemond/software/radmc-3d/download.html}.}. In addition, sharp increase of phase function makes opacity integration inaccurate. In order to avoid these problems, we place upper limit on peaking forward scattering, since forward scattering outside the observable angle range is not likely to affect observed images. In this paper, we adopt $Z_{11}(\theta<\theta_c)=Z_{11}(\theta_c)$, where $\theta_c$ is taken to be 1 degree.

\begin{figure*}
\begin{center}
\includegraphics[height=6.0cm,keepaspectratio]{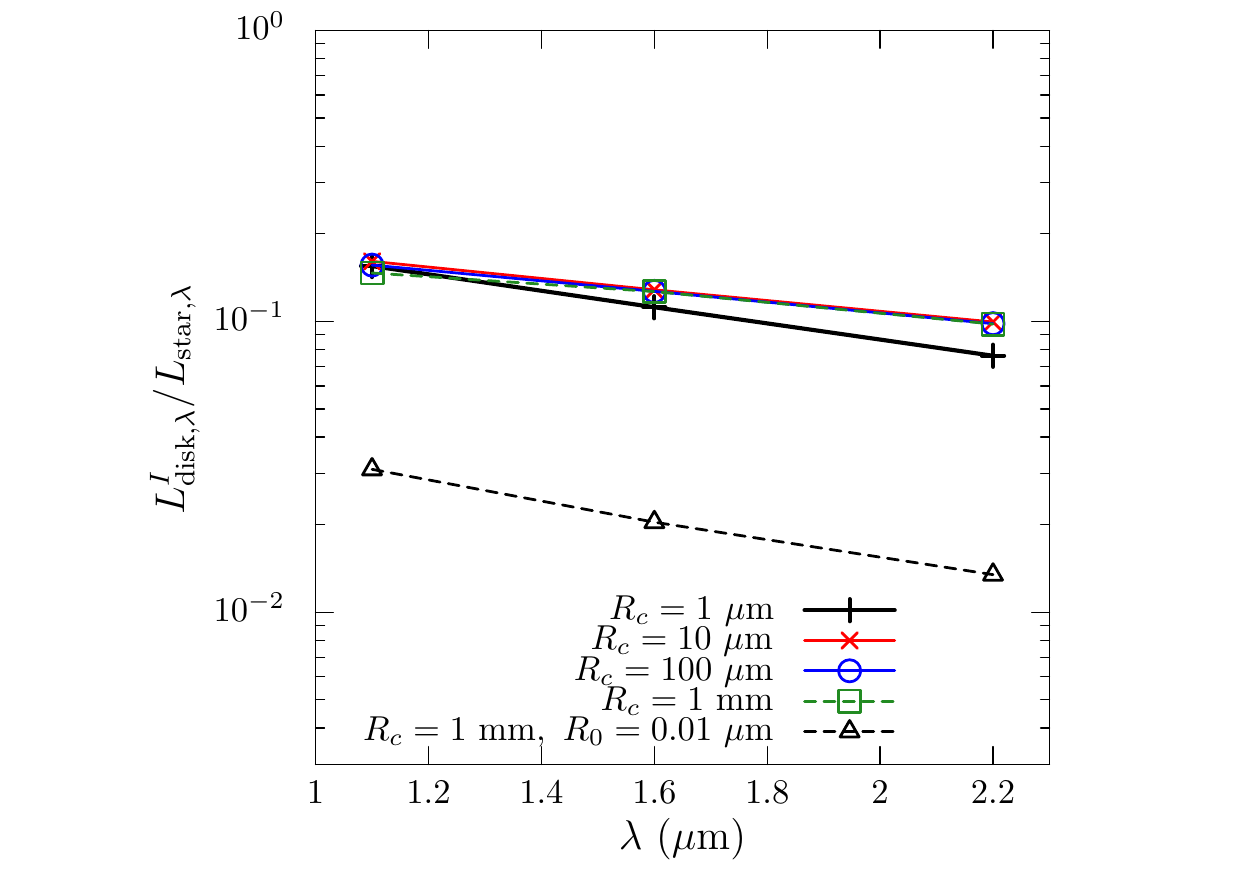}
\includegraphics[height=6.0cm,keepaspectratio]{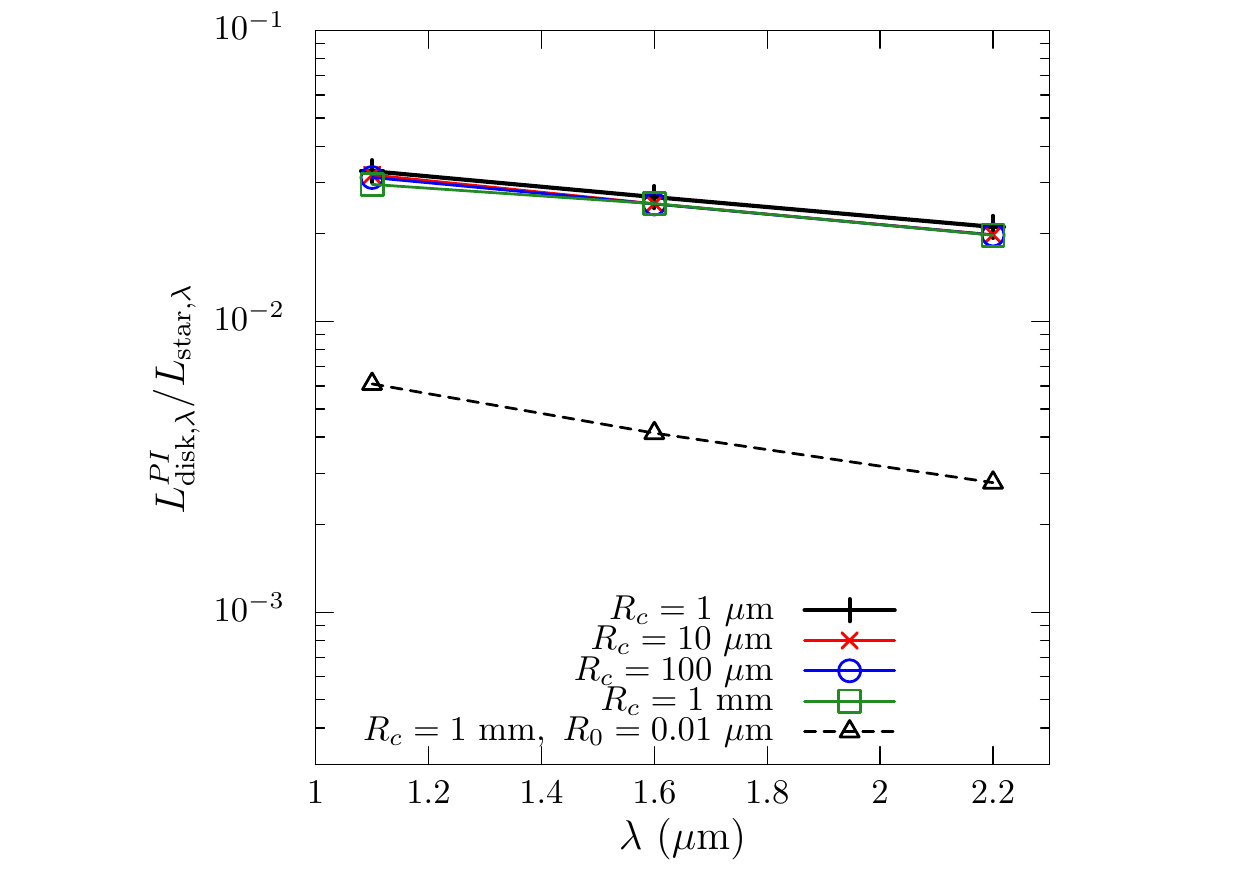}
\caption{Fractional luminosity of the disc for total intensity (left) and polarised intensity (right).}
\label{fig:large_colour}
\end{center}
\end{figure*}

Figure \ref{fig:large_colour} shows results of radiative transfer simulations of discs containing BCCA with $R_c=1\ \mu$m to $1$ mm (Obtained images are shown in Figure \ref{fig:extlargeimage}.).
It is found that the disc scattered light is insensitive to the characteristic radius of aggregate. This is because scattered light intensity in a range of observable scattering angles does not depend on the characteristic radius (see hatched range of scattering angles in Figure \ref{fig:Z11large}). Table \ref{table:large_colour} shows that colours become shallower as the characteristic radius increases; however, even if the aggregate radius is increased up to millimetre-size, colours do not reach the reddish colour regime. Colours of BCCA with $R_0=0.1\mu$m seem to be slightly blue, but not too blue like Rayleigh scattering particles (see Table \ref{table:colour_porous}). Therefore, even if the disc contains BCCA as large as millimetre-size at the surface, reddish colour scattered light is not likely to occur.

In Figure \ref{fig:Z11large}, we also investigate how the monomer radius affects disc scattered lights. In order to satisfy Equation (\ref{eq:cond1}), we consider millimetre-sized BCCA with $R_0=0.01\ \mu$m-sized monomers. If the monomer radius is decreased, the disc becomes faint due to small albedo value of dust aggregates. Large BCCA of small monomers show bluer colours than those of $R_0=0.1\ \mu$m monomers. Therefore, scattered light colours of BCCA depend on the monomer radius (see Section \ref{sec:why} for more detail).

Finally, we mention error of scattered light colours due to the use of MMF (see also Appendix \ref{sec:approx}).  
MMF results show slightly blue colours for large BCCA; however, MMF tends to produce bluer colours compared to rigorous TMM calculations (see Table \ref{table:large_colour}). For $R_c=4.8\ \mu$m, the difference of the slope is $\Delta_\mathrm{I}\equiv\eta_\mathrm{I, TMM}-\eta_\mathrm{I, MMF}\simeq0.16$, where $\eta_\mathrm{I, TMM}$ and $\eta_\mathrm{I, MMF}$ are total intensity colours obtained by TMM and MMF, respectively. 
Similarly, we can define the difference of the slope for polarised intensity, $\Delta_\mathrm{PI}
\equiv\eta_\mathrm{PI, TMM}-\eta_\mathrm{PI, MMF}\simeq0.26$. Hence, TMM results are shallower than MMF results (see Figure \ref{fig:colourincl0}). These difference are mainly due to approach of $\Delta\phi$ to unity, which makes MMF inaccurate due to occurrence of multiple scattering.
$\Delta\phi$ increases as the wavelength decreases; however, for sufficiently large BCCA (e.g., $d_f\approx2$), $\Delta\phi$ does not depends on the aggregate radius and depends only on the property of the monomer (refractive index and size parameter) \citep{Berry86, Tazaki18}. 
Hence, it is reasonable to assume that errors in slope for $R_c=10\ \mu$m, 100$\ \mu$m, and $1$ mm are similar to those of $R_c=4.8 \mu$m, and therefore, slope of these large BCCA may be shallower than those given in Table \ref{table:large_colour} by $\Delta_\mathrm{I}=0.16$ and $\Delta_\mathrm{PI}=0.26$ for total intensity and polarised intensity, respectively. 
As a result, by considering errors in colours between TMM and MMF, millimetre-sized BCCA with $R_0=0.1\ \mu$m are thought not to produce reddish colours at near-infrared wavelengths and more likely to produce grey colours.

\begin{table}
  \caption{Disc Scattered Light Colour for extremely large BCCA}
  \label{table:large_colour}
  \centering
  \begin{tabular}{lcc}
    \hline
    Aggregate Radius  & $\eta_\mathrm{I}$ & $\eta_\mathrm{PI}$ \\
    \hline \hline
    $R_c=1\ \mu$m (MMF)  & $-1.0$ & $-0.64$  \\
    $R_c=10\ \mu$m (MMF)  & $-0.69$ & $-0.68$  \\
    $R_c=100\ \mu$m (MMF)  & $-0.67$ & $-0.66$  \\
    $R_c=1$ mm (MMF)  & $-0.58$ & $-0.58$ \\
    $R_c=1$ mm, $R_0=0.01\ \mu$m (MMF)  & $-1.2$ & $-1.1$ \\
    \hline
    $R_c=4.8\ \mu$m (TMM)  & $-0.51$ & $-0.41$ \\
    $R_c=4.8\ \mu$m (MMF)  & $-0.67$ & $-0.67$ \\
    \hline
 \end{tabular}
\end{table}

\subsection{Effect of disc geometry, inclination, and dust composition} \label{sec:gene}
\begin{figure}[t]
\begin{center}
\includegraphics[height=6.0cm,keepaspectratio]{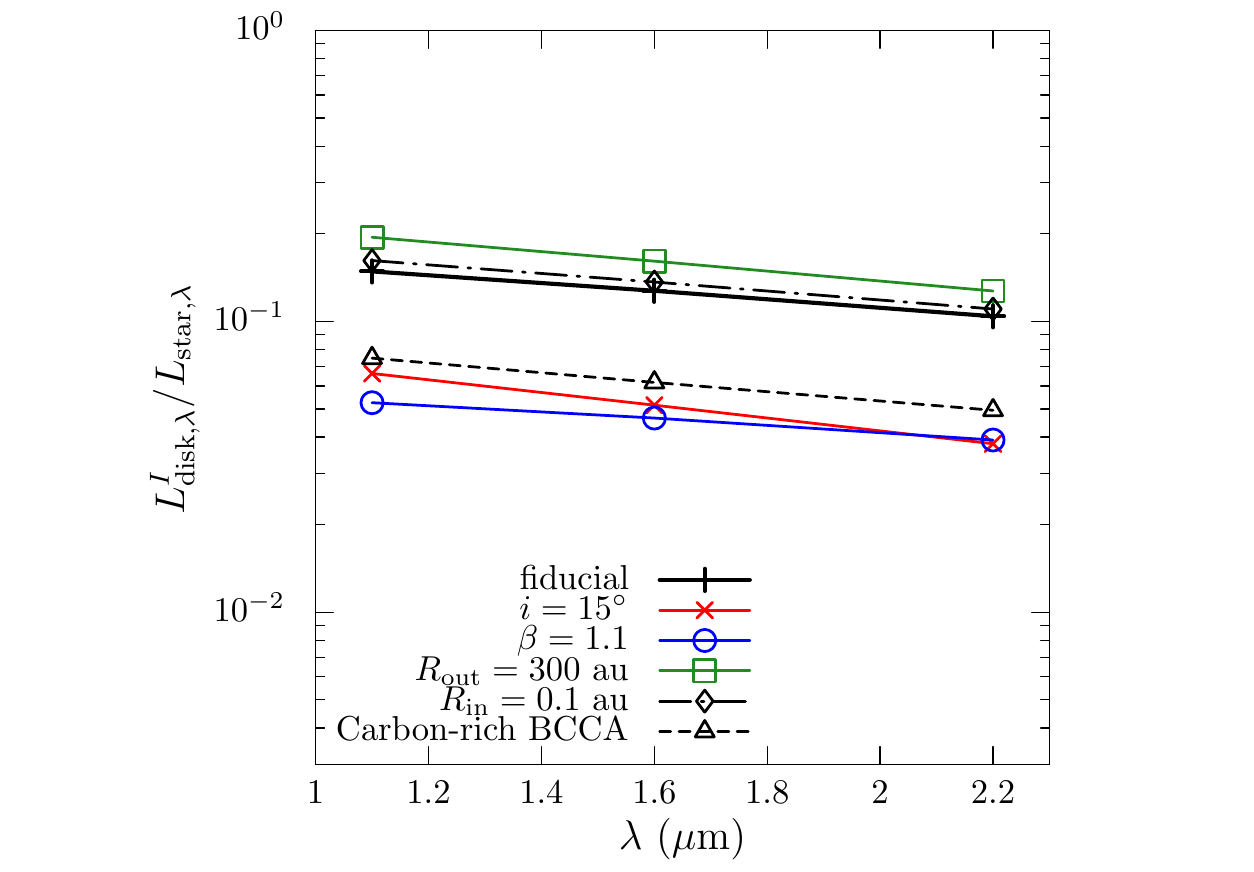}
\caption{Fractional luminosity for total intensity for various disc and dust composition, where the dust morphology is the BCCA model ($R_c=4.8\ \mu$m and $R_0=0.1\ \mu$m). The fiducial model has the inclination angle $i=60^\circ$, the flaring index $\beta=1.25$, the inner and outer radii $R_\mathrm{in}=10$ au and $R_\mathrm{out}=100$ au, respectively, and the dust composition is astronomical silicate (plus symbol). Each parameter is changed one-by-one from the fiducial model; the fiducial model, but for $i=15^\circ$ (cross), for $\beta=1.1$ (circle), for $R_\mathrm{out}=300$ au (square), for $R_\mathrm{in}=0.1$ au (diamond), and for carbon-rich BCCA (triangle).}
\label{fig:modeldep}
\end{center}
\end{figure}

In Sections \ref{sec:compporous} and \ref{sec:exlarge}, we studied scattered-light colour by assuming a single disc geometry, inclination angle, and a single dust composition. Here, we address how these parameters affect scattered light colours.

We vary following parameters: flaring index $\beta$, disc inner and outer truncation radius $R_\mathrm{in}$ and $R_\mathrm{out}$, respectively, inclination angle $i$, and dust composition. Total dust mass is kept the same, that is $10^{-4}M_\odot$.
As a dust composition, we consider the carbon-rich composition, which is a mixture of silicate, iron sulphide, water ice, and amorphous carbon. 
Fraction of each ingredients are determined by a recipe given by \citet{Min11} with the carbon partition parameter $w=1$. Optical constants of silicate, iron sulphide, and water ice are taken from \citet{Henning96} and for amorphous carbon from \citet{Zubko96}. These optical constants are mixed by using the Bruggeman mixing rule \citep{Bruggeman1935}. Optical properties of carbon-rich BCCA ($R_c=4.8\ \mu$m and $R_0=0.1\ \mu$m) are obtained by using MMF. 

In Figure \ref{fig:modeldep}, we show scattered-light colour of the BCCA model for various parameters. Table \ref{table:colour_model} summarises scattered-light colours. It is shown that that these parameters affect mainly magnitude of fractal luminosity. 
Although colours can differ for different disc geometries and dust composition, within a parameter range we studied, these parameters do not significantly change the our results given in Sections \ref{sec:compporous} and \ref{sec:exlarge}.

It should be mentioned that fractional luminosity in total intensity and polarised intensity derived in this paper is higher than the observed values \citep{Fukagawa10, Avenhaus18}. However, as shown in Figure \ref{fig:modeldep}, disc scattered light luminosity is sensitive to the disc structure and dust composition, although the colours are less sensitive to them. Hence, we expect observed fractional luminosity in total intensity and polarised intensity might be explained by including these effects, though we need to check it for each object. This is beyond the scope of this paper.

\begin{table}
  \caption{Disc Scattered Light Colour for Different Disc and Dust Models}
  \label{table:colour_model}
  \centering
  \begin{tabular}{lcc}
    \hline
    Dust model  & $\eta_\mathrm{I}$ & $\eta_\mathrm{PI}$ \\
    \hline \hline
    fiducial  & $-0.51$ & $-0.41$ \\
    $i=15^\circ$ & $-0.80$ & $-0.46$ \\
    $\beta=1.1$ & $-0.43$ & $-0.30$ \\
    $R_\mathrm{out}=300$ au & $-0.61$ & $-0.47$ \\ 
    $R_\mathrm{in}=0.1$ au & $-0.55$ & $-0.47$ \\ 
    Carbon-rich BCCA & $-0.60$ & $-0.66$ \\   
    \hline
 \end{tabular}
\end{table}

\section{Effect of dust aggregate structure on scattering property} \label{sec:why}
\begin{figure*}
\begin{center}
\includegraphics[height=6.0cm,keepaspectratio]{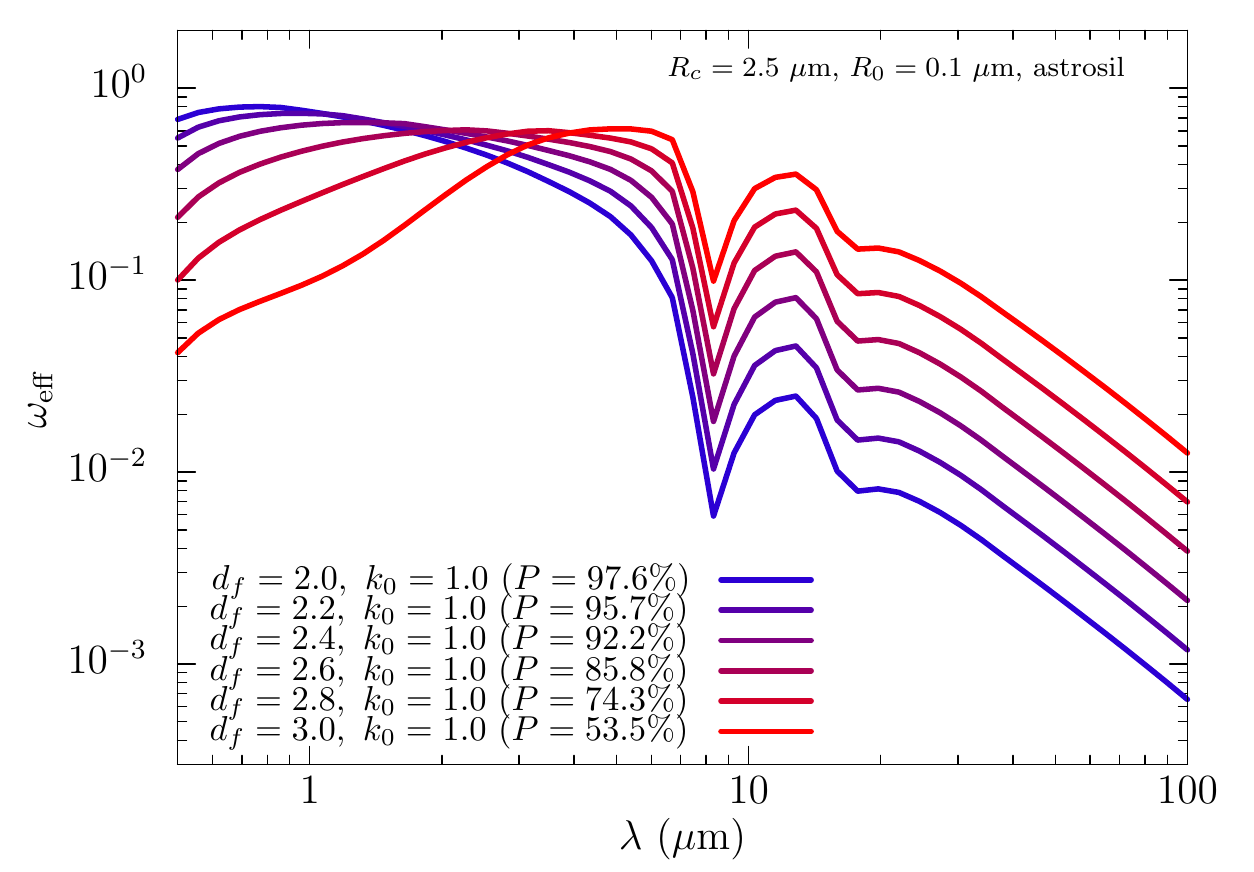}
\includegraphics[height=6.0cm,keepaspectratio]{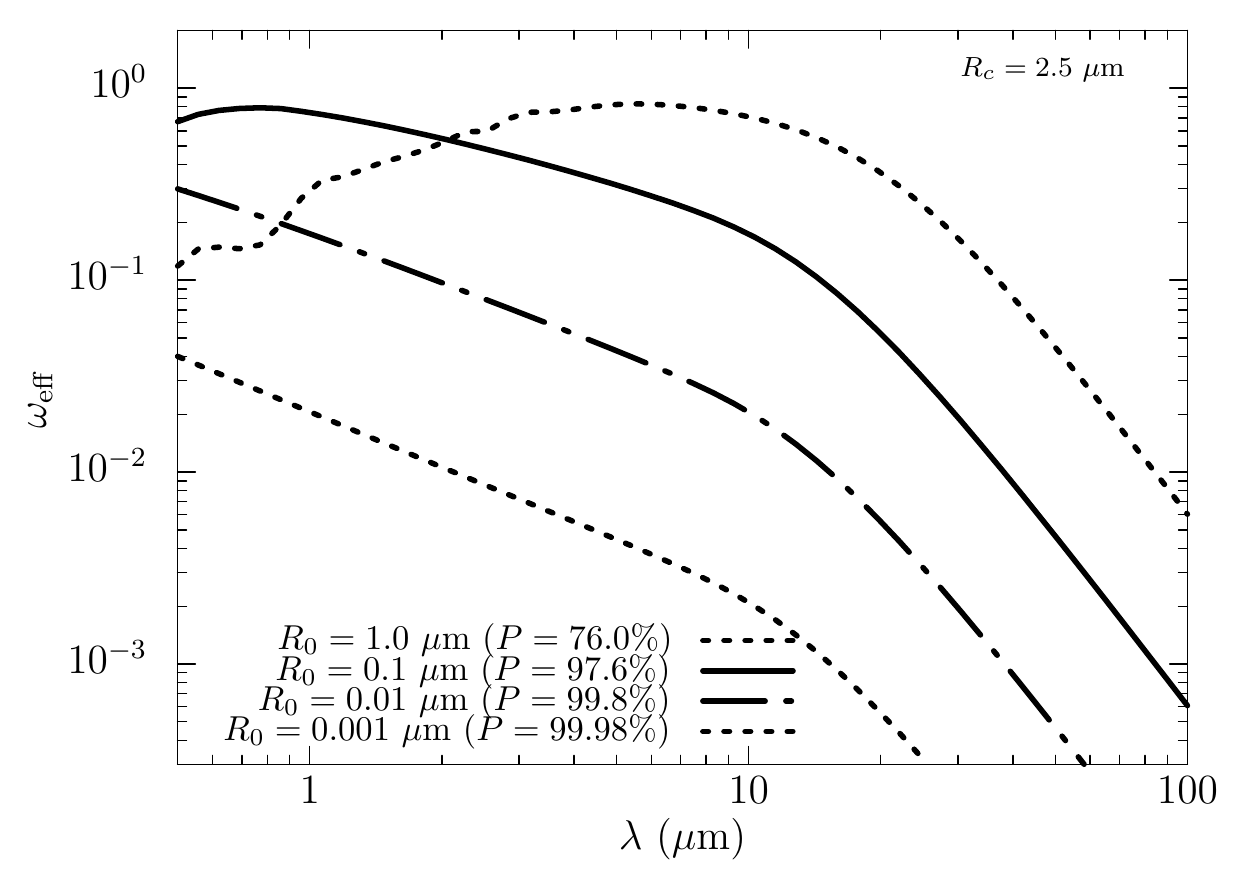}
\caption{(Left) The effective albedo for different values of $d_f$ and $k_0$, where the characteristic radius and monomer radii are $R_c=2.5\ \mu$m and $R_0=0.1\ \mu$m, respectively. Refractive indices is astronomical silicate. 
(Right) The effective albedo of fluffy dust aggregates ($k_0=1.0$ and $d_f=2$) of various monomer radii. The characteristic radius is assumed to be 2.5 $\mu$m. Refractive indices is set as $m=1.67+0.0326i$, which corresponds to those of astronomical silicate at $\lambda=1.6\ \mu$m.}
\label{fig:ealbd_ccpc}
\end{center}
\end{figure*}

In Section \ref{sec:colour}, it is shown that large porous dust aggregates show slightly blue or grey colours in scattered light, while large compact aggregates tends to show reddish colours. In this section, we discuss reasons for this by considering intrinsic optical properties of dust aggregates. 

We define the effective albedo by $\omega_\mathrm{eff}=\kappa_\mathrm{sca}^\mathrm{eff}/(\kappa_\mathrm{abs}+\kappa_\mathrm{sca}^\mathrm{eff})$,
where $\kappa_\mathrm{sca}^\mathrm{eff}=(1-g)\kappa_\mathrm{sca}$ is the effective scattering opacity, $\kappa_\mathrm{abs}$ is the absorption opacity, and $g$ is the asymmetry parameter. Since small angle scattered light is hardly observable, the effective albedo is a useful quantities \citep[see also][]{Dullemond03, Min10, Mulders13}. In the case of efficient forward scattering ($g\sim1$), only small amount of incident photons will be scattered toward the observer (supposed to be at $\theta\neq0$); hence, the effective albedo has small values. On the other hand, if scattering is isotropic ($g\sim0$), the effective albedo gives rise to the single scattering albedo $\omega$. Thus, the effective albedo might be used as a qualitative measure of disc scattered light colours. 

In order to obtain the effective albedo of dust aggregates, we use MMF developed by \citet{Tazaki18}. In MMF, the structure of dust aggregates is specified in terms of two-point correlation function of monomers \citep{Berry86, Botet97, Tazaki18}. In addition, by using of mean field assumption, multiple scattering inside the aggregate is solved in a self-consistent manner. Hence, it is suitable to study how fractal dimension affect optical cross sections.

Figure \ref{fig:ealbd_ccpc} shows the effective albedo for dust aggregates with various fractal dimension and monomer radii. It is found that when the characteristic radius is fixed to 2.5 $\mu$m, with increasing fractal dimension from 2 to 3, colours of the effective albedo vary from blue to red at near-infrared wavelengths. 

This can be explained by as follows. For large compact dust aggregates, as wavelength decreases, forward scattering becomes more efficient, that is, $g$ increases. Thus, the effective albedo decreases for short wavelength domain \citep{Mulders13}. 
Hence, large compact aggregates shows reddish colours.
Meanwhile, for large highly porous aggregates, asymmetry parameter $g$ becomes almost constant with respect to the wavelength due to saturation of scattered light (see Appendix \ref{sec:AppendixC}), and therefore, anisotropic scattering does not reduce the effective albedo. As a result, porous aggregates tends to show more bluer colours compared to compact aggregates.

Scattered light colours depend also on the monomer radius, in particular for $d_f=2$ (like BCCA). In right panel of Figure \ref{fig:ealbd_ccpc}, we show the effective albedo of fluffy dust aggregates ($k_0=1.0$ and $d_f=2.0$) with various monomer radii, where the complex refractive indices is set as $m=1.67+0.0326i$ for the sake of simplicity (corresponding to astronomical silicate at $\lambda=1.6\ \mu$m). It is found that when the monomer radius is much smaller that the NIR wavelength (cases of $R_0=1$ nm and 10 nm), NIR slope of the effective albedo is blue (see also Table \ref{table:large_colour}). On the other hand, for the case of $R_0=1\ \mu$m, the single scattering assumption is violated, and hence, the effective albedo becomes red due to occurrence of multiple scattering inside the cluster. Thus, the monomer radius is sensitive to scattered light colours of large fluffy dust aggregates. 

There are three possibilities to explain observed grey scattered light colours. 
One possibility is moderately compressed aggregates. 
As shown in Figure \ref{fig:ealbd_ccpc}, scattered light colours of aggregates with $d_f\approx2.5$ will be almost grey. Numerical simulations have shown that mutual aggregate collisions can only produces an aggregate with fractal dimension $d_f\approx2.5$ \citep{Wada08, Suyama08}. Hence, collision compressed aggregates may produce grey scattered light colours at near-infrared wavelength. Another possibility is that the aggregate with $d_f\approx2$ has the monomer radius slightly larger than 0.1 $\mu$m but not as large as micron size. This can also produce grey colours. The third possibility is the conventional one, that is, compact particles with the radius comparable to the observing wavelength (Figure \ref{fig:DHScolour}). Although composite mixture of compact particles and porous aggregates in the same disk may also affect disk colours, this is left for a future task.

\section{Comparison with disc observations} \label{sec:implication}
Near-infrared observations of protoplanetary discs have shown that most discs show grey colours in both total and polarised intensity \citep{Fukagawa10, Avenhaus18}. Previously, these scattered-light properties are interpreted as the presence of compact grains with the radius comparable to the observing wavelength (see Figure \ref{fig:DHScolour}). 
In this case, it is necessary that dust grain radii at disc surface is adjusted to observing wavelength. One adjusting mechanism could be dust vertical stratification \citep{Duchene04, Pinte07}. This study provides a new insight into interpretation of discs showing grey colours. We found that large porous dust aggregates can also show marginally grey or slightly blue total intensity. One advantage is that in the case of porous aggregates, their radius is not necessary to adjust to observing wavelength. In addition, large porous dust aggregates are expected to be stirred higher altitude of the discs, they are likely to affect disc scattered light. Therefore, grey discs in total intensity might by explained by large fluffy dust aggregates, although more detail modelling is necessary for each object.

If a disc contains porous dust aggregates, we predict that polarisation fraction of disk scattered light should be high. For example, as shown in Figure \ref{fig:polfrac}, porous aggregates with sub-micron monomers produces polarisation fraction as high as 65\%-75\% and disc integrated polarisation fraction is about 18\% at $\lambda=1.6\ \mu$m at inclination angle $i=60^\circ$  (see also Table \ref{table:pfrac}). 

Spatially resolved polarisation fraction map is obtained for several discs, and these observations have revealed that disc scattered light in near-infrared wavelengths is often highly polarised. Polarisation fraction of GG Tau and AB Aur discs show polarisation fraction as high as 50\% at $\lambda=1\ \mu$m for GG Tau \citep{Silber00} and $\lambda=2\ \mu$m for AB Aur \citep{Perrin09}.
More recently, \citet{Itoh14} reported polarisation fraction of GG Tau is as high as that of Rayleigh scattering function at H-band; nevertheless, strong forward scattering is also observed at H-band. For HD 142527, \citet{Canovas13} derived polarisation fraction of 10\%--25\% at H-band; however, \citet{Avenhaus14} reported significantly higher polarisation fraction for this object (20\%--45\%) at H-band.
\citet{Tanii12} show that polarisation fraction of UX Tau is up to 66\% at H-band. 
Furthermore, \citet{Poteet18} have shown that polarisation fraction of TW Hya is as high as $63\%\pm9\%$. 
Thus, scattered light of these spatially resolved objects is highly polarised, and hence these scattered light might be explained by porous dust aggregates models, although both detail modelling for each object and further observational studies to derive polarisation fraction are necessary. It should be emphasised that polarisation fraction predicted by our porous dust aggregate model is not {\it very} high compared to disc observations. 

\citet{Mulders13} pointed out that some protoplanetary discs show reddish scattered-light colour in total intensity. For the case of the outer disc of HD 100546, reddish and faint scattered light are observed, and it can be explained by 2.5 $\mu$m sized spherical particles \citep{Mulders13}. Based on the SED modelling of HD 100546, \citet{Mulders13} also claimed that the disc scale height is consistent with a particle model of 0.1 $\mu$m rather than 2.5 $\mu$m for a given turbulent strength and disc gas mass. They suggested a possibility that the presence of {\it extremely fluffy} dust aggregates containing submicron-sized monomers at the disc surface layer may reconcile this apparent conflict, since such dust aggregate behaves like a small particle in dynamics and like a large particle in light scattering process. However, we found that fluffy aggregates of submicron-sized monomers do not show reddish colour in total intensity. Reddish scattered light is more likely to be observed when the dust aggregates have compact structure (see Figures \ref{fig:DHScolour} and \ref{fig:ealbd_ccpc}).
Therefore, our results imply that fluffy dust aggregates are not responsible for the scattered-light properties of the outer disk of HD 100546. 
Our results predict that polarisation fraction of HD 100546 should not be high.
\citet{Quanz11} show integrated polarisation fraction of this disc is about $14\%$.
Since average inclination angle of HD 100546 is $46^\circ$ \citep{Mulders13}, observed polarisation fraction is lower than the prediction of our BCCA model, which shows about 32\% at $i=45^\circ$ (Table \ref{table:pfrac}). This observed polarisation fraction is more close to our compact dust aggregate model. Recently, \citet{Stolker16b} derived scattering phase function of this object. The derived phase function increases with increasing scattering angles from intermediate to back scattering angles. This enhanced backward scattering can be seen in phase function of compact dust aggregates \citep{Min16}, and not for fluffy dust aggregate \citep[][see also Figure \ref{fig:porousring}]{Tazaki16}. Laboratory experiments also support that large compact particles show the enhanced backscattering \citep{Munoz17}. These enhanced backscattering presumably due to the presence of wavelength scale surface roughness of a large compact particle \citep[e.g.,][]{Mukai82}. Thus, phase function of this object is consistent with our implication. 

\section{Conclusion} \label{sec:conclusion}
We have studied how radius and structure of dust aggregates affect observational quantities of protoplanetary discs in near-infrared wavelengths. We have performed radiative transfer calculations of protoplanetary discs taking fluffy dust aggregates into account, where we firstly treated their optical properties in a proper manner. Furthermore, based on an approximate theory for optical properties of fractal dust aggregates (MMF) \citep{Tazaki16, Tazaki18}, we have argued scattering properties of fractal dust aggregates. Our primary findings are summarised as follows.

\begin{itemize}
\item The ratio of a aggregate radius and wavelength can be assessed by the presence of brightness asymmetry in total intensity images. Both porous and compact dust aggregates can produce brightness asymmetry in total intensity when their radii exceed the observing wavelength (Sections \ref{sec:dustoptprop} and \ref{sec:image}). 

\item Polarisation fraction of a disc is useful to probe structure of dust aggregates 
as long as the aggregate radius exceeds observing wavelength (Sections \ref{sec:dustoptprop} and \ref{sec:image}). Higher porosity produces higher polarisation fraction. We have provided expected integrated polarisation fraction for various dust models and disk inclinations (Table \ref{table:pfrac}).

\item Porous dust aggregates (BCCAs and BPCAs) large compared to near-infrared wavelengths show marginally grey or slightly blue in total or polarised intensity. Meanwhile, large compact dust aggregates show reddish scattered light colour in total intensity (Section \ref{sec:compporous}). 

\item For sufficiently large BCCA, the aggregate radius is less sensitive to disc scattered light because of saturation of scattering property (Figures \ref{fig:Z11large} and \ref{fig:saturation}). Even if the radius is increased up to millimetre-size, BCCA containing $0.1\ \mu$m-sized monomers is expected to show marginally grey or slightly blue colours (Section \ref{sec:exlarge}). 

\item Within a parameter range we studied, the disc geometry, inclination angle, and composition seems not to be sensitive to the disc scattered light colours in near-infrared wavelengths, although these can affect magnitude of fractional luminosity (Section \ref{sec:gene}).

\item The effective albedo for aggregate with various fractal dimensions and monomer radius are computed. As fractal dimension or monomer radius increases, wavelength dependence of the effective albedo varies from blue to red at near-infrared wavelengths. BCCA can show reddish colours in near-infrared wavelengths when the monomer is about micron-sized. On the other hand, smaller monomers makes BCCA a bluer scatterer in near-infrared wavelength (Section \ref{sec:why}).
\end{itemize}

\section*{Acknowledgements}
We sincerely thank the referee for a thorough and careful reading of the manuscript.
R.T. would like to thank Daniel Mackowski and Yasuhiko Okada for the availability of the T-Matrix code with the QMC method. R.T. also thanks Cornelis P. Dullemond for making the \texttt{RADMC-3D} code public. R.T. thanks Robert Botet for fruitful discussion. R.T. was supported by a Research Fellowship for Young Scientists from the Japan Society for the Promotion of Science (JSPS) (17J02411).




\bibliographystyle{mn2e}



\appendix
\section{Comparison between approximate methods} \label{sec:approx}
We perform radiative transfer simulation of the BCCA model with approximate methods; modified mean field theory \citep[MMF;][]{Tazaki18}, effective medium theory with Maxwell Garnett mixing rule \citep[EMT;][]{Mukai92, Kataoka14}, and DHS, and then the results are compared to those obtained by TMM.

\begin{figure*}
\begin{center}
\includegraphics[height=13.0cm,keepaspectratio]{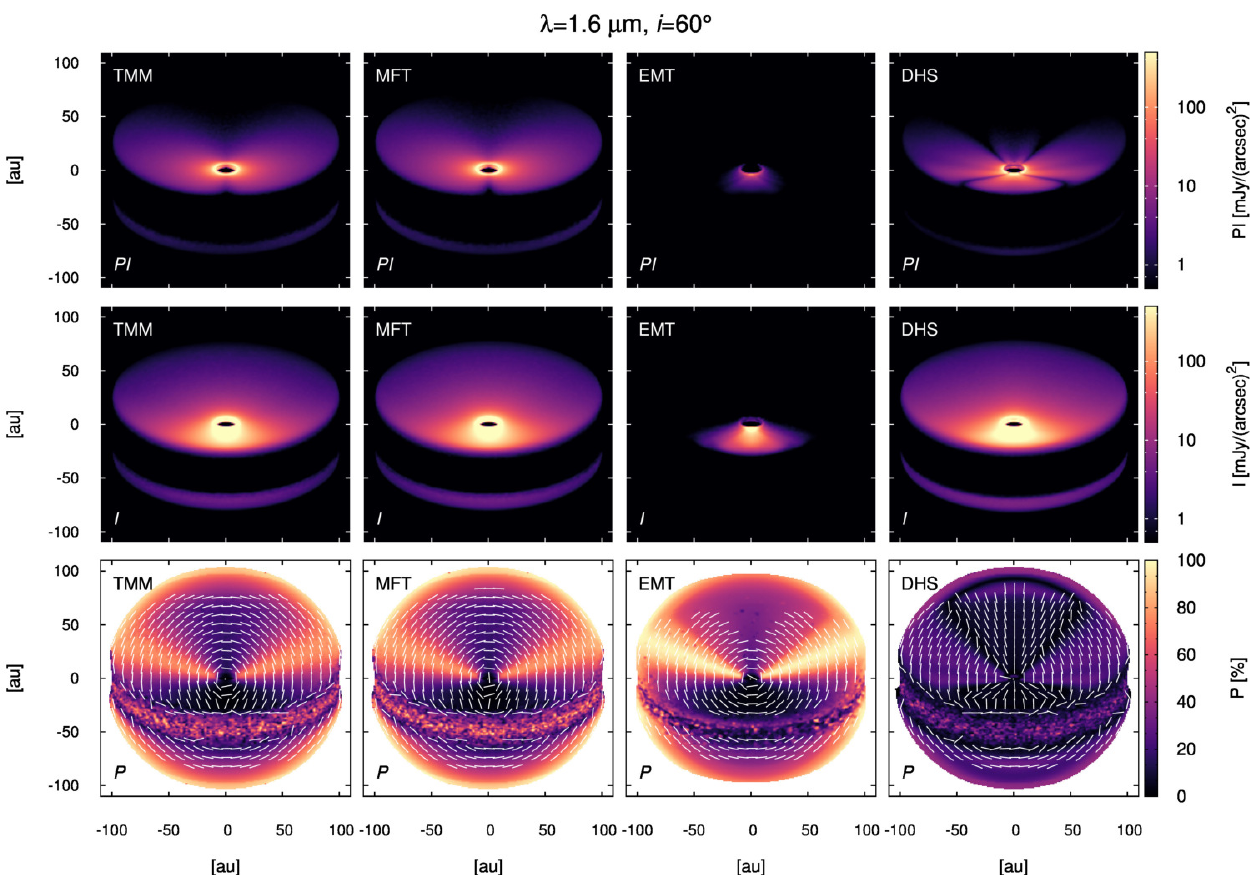}
\caption{Polarised intensity (top), total intensity (middle), and polarisation fraction (bottom). From left to right, TMM, MFT, EMT, and DHS are used. Wavelength is $\lambda=1.6\ \mu$m and the inclination angle is $i=60^\circ$. The dust model adopted here is BCCA with $N=1024$ and $R_0=0.1\ \mu$m, corresponding to the characteristic radius $R_c=4.8\ \mu$m.}
\label{fig:rtincl60}
\end{center}
\end{figure*}

\begin{figure*}
\begin{center}
\includegraphics[height=6.0cm,keepaspectratio]{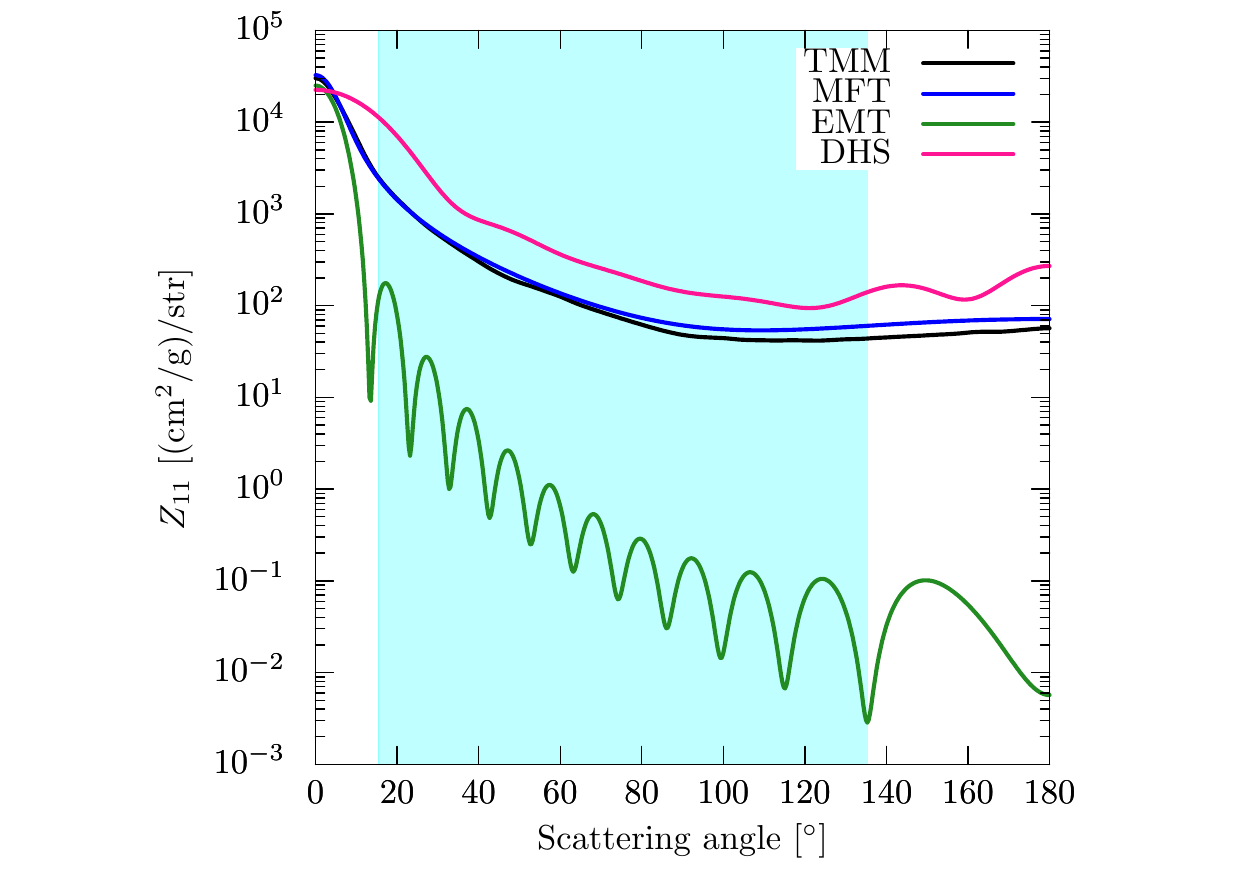}
\includegraphics[height=6.0cm,keepaspectratio]{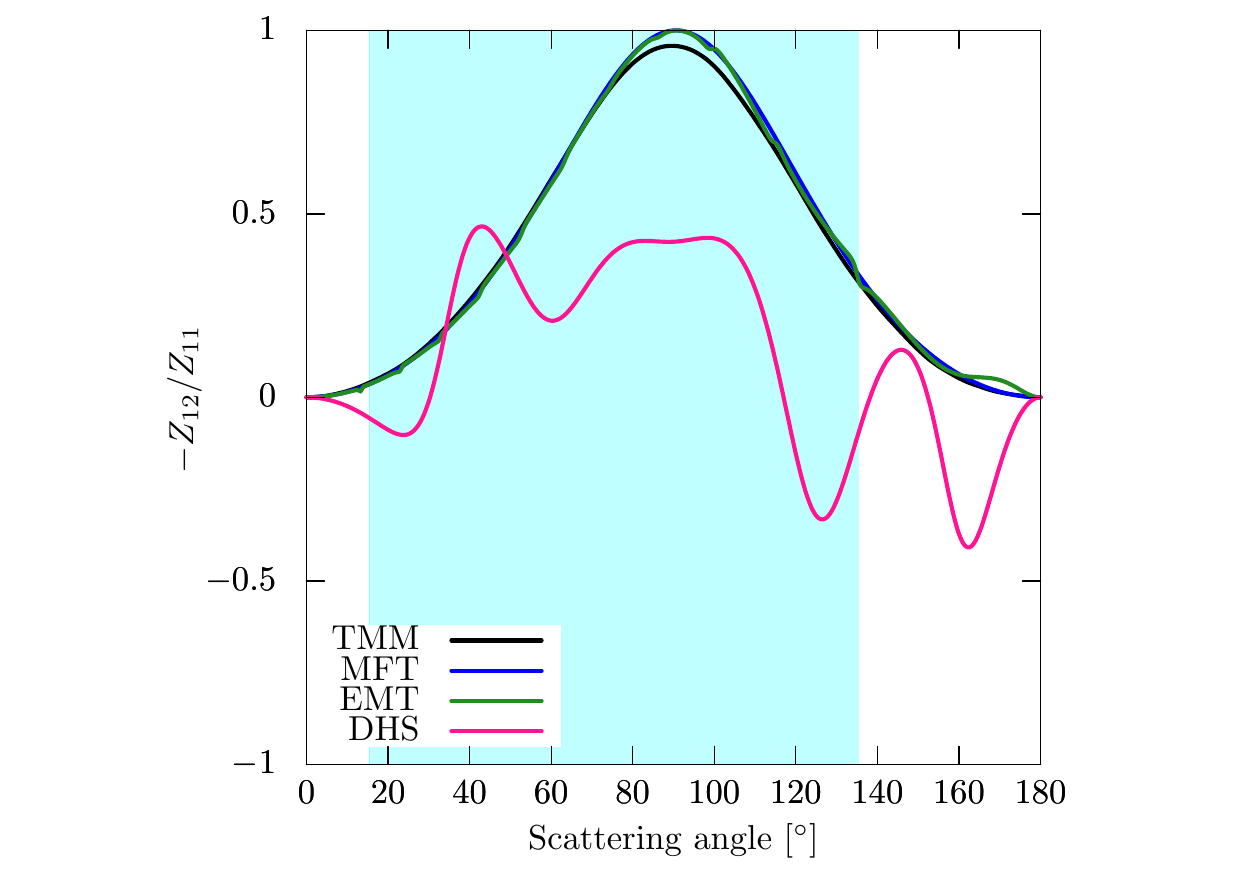}
\caption{Phase function $Z_{11}$ (left), and the degree of linear polarisation $-Z_{12}/Z_{11}$ (right). The dust model is the BCCA model. Black line is the result obtained by TMM method (rigorous numerical method), while others (MFT, EMT, DHS) are those obtained by using the approximate method for optical property calculations. MFT can reproduce the TMM results for both intensity and polarisation fraction, whereas EMT and DHS fail. Hatched region indicates a range of scattering angle to be observed for a disc with the flaring index $\beta=1.25$ and the inclination angle $i=60^\circ$.}
\label{fig:ring}
\end{center}
\end{figure*}

\begin{figure}
\begin{center}
\includegraphics[height=6.0cm,keepaspectratio]{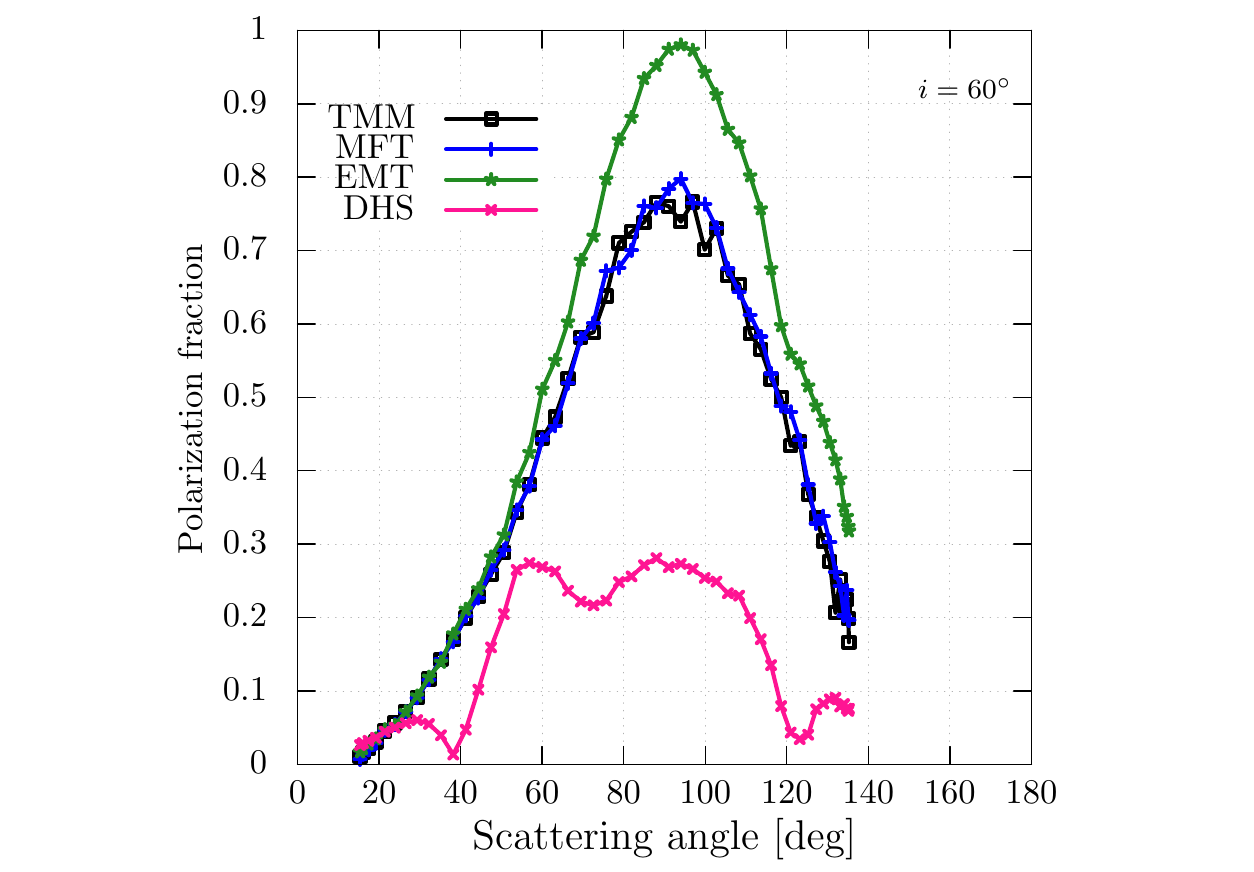}
\caption{Polarisation fraction as a function of scattering angles at $R=50$ au measured from simulated images in Figure \ref{fig:rtincl60}.}
\label{fig:polfrac_approx}
\end{center}
\end{figure}

\begin{table}
  \caption{Disc Scattered Light Colour for BCCA with Different Methods}
  \label{table:colour_method}
  \centering
  \begin{tabular}{lcc}
    \hline
    Method  & $\eta_\mathrm{I}$ & $\eta_\mathrm{PI}$ \\
    \hline \hline
    TMM  & $-0.51$ & $-0.41$ \\
    MFT  & $-0.67$ & $-0.67$ \\
    EMT  & $0.53$ & $0.65$ \\
    DHS  & $0.030$ & $0.099$ \\
    \hline
 \end{tabular}
\end{table}

Figure \ref{fig:rtincl60} shows scattered-light images of the disc at the wavelength of $\lambda=1.6\ \mu$m and the inclination angle of $i=60^\circ$ where the dust model is the BCCA model with $N=1024$ and $R_0=0.1\ \mu$m. As shown in \citet{Tazaki18}, for this BCCA model, MMF coincide with the mean field theory (MFT) results at $\lambda>0.85\ \mu$m. Hence, we use MFT instead of MMF.

Figure \ref{fig:rtincl60} clearly shows that MFT reproduces the TMM results, whereas EMT and DHS fail.

First of all, we study the difference of each method in total intensity of the discs (middle panels in Figure \ref{fig:rtincl60}). All methods show the front-back asymmetry.
Both MFT and DHS show similar results to the TMM result; however, the EMT result significantly deviates from the TMM result. This is because EMT produces extremely faint backward scattering (Figure \ref{fig:ring}, see also Figure 5 of \citet{Tazaki16}) due to destructive interference of scattered waves. In other words, extremely strong forward scattering predicted by the EMT model means that most of the photons coming from the star are just passing through the particle, and they are not likely to be scattered toward the observer. As a result, when EMT is applied to obtain optical properties of BCCA, the disc scattered light becomes too faint. 

Secondly, we study polarisation fraction (bottom panels in Figure \ref{fig:rtincl60}). 
Figure \ref{fig:rtincl60} shows that TMM and MFT show almost the same polarisation fraction; however, EMT and DHS overestimates and underestimates polarisation fraction, respectively (Figure \ref{fig:polfrac_approx}).
The low polarisation fraction of DHS is due to scattering property of the particle (Figure \ref{fig:ring}). The degree of polarisation of the BCCA model at wavelength $1.6\ \mu$m shows has a bell shaped angular profile (almost symmetric with respect to $\theta=90^\circ$), and maximum degree of polarisation is 96\% at $\theta=90^\circ$ (Figure \ref{fig:ring}, see also Figure 15 of \citet{Tazaki16}). However, the DHS model predicts the degree of polarisation is 40\% at $\theta=90^\circ$, which is significantly lower than the TMM result. As a consequence, polarisation fraction of DHS becomes lower than that of TMM.
The overestimation of polarisation fraction of EMT is due to a radiative transfer effect. 
As shown in \citet{Tazaki16}, both EMT and MFT show a similar polarisation profile to TMM \footnote{The degree of polarisation obtained by MFT is almost the same as that of the RGD theory because phase shift is less than unity for the dust model and wavelength now we concern (see also \citet{Tazaki18}).}. Nevertheless, EMT significantly overestimates polarisation fraction. As we have mentioned, in the EMT model, large angle scattering is suppressed by destructive interference, and hence, multiple scattering at the disc surface layer is not likely to occur. Since MFT can reproduce phase function obtained by the TMM correctly (see Figure 5 in \citet{Tazaki16}), MFT can mimic the depolarisation effect in radiative transfer, whereas EMT fails.

The DHS model show some slits in the image of polarised intensity. This is because oscillatory behaviour in a angular profile of the degree of polarisation originated from resonance of smooth spherical surface. Since the degree of polarisation becomes positive and negative in the oscillatory behaviour, polarisation vectors in DHS are very different from the TMM results. In the DHS model, optical properties are averaged over the size distribution; nevertheless, the oscillatory behaviour remains. This is mainly because astronomical silicate is less absorbing at infrared wavelengths. 

\begin{figure*}
\begin{center}
\includegraphics[height=6.0cm,keepaspectratio]{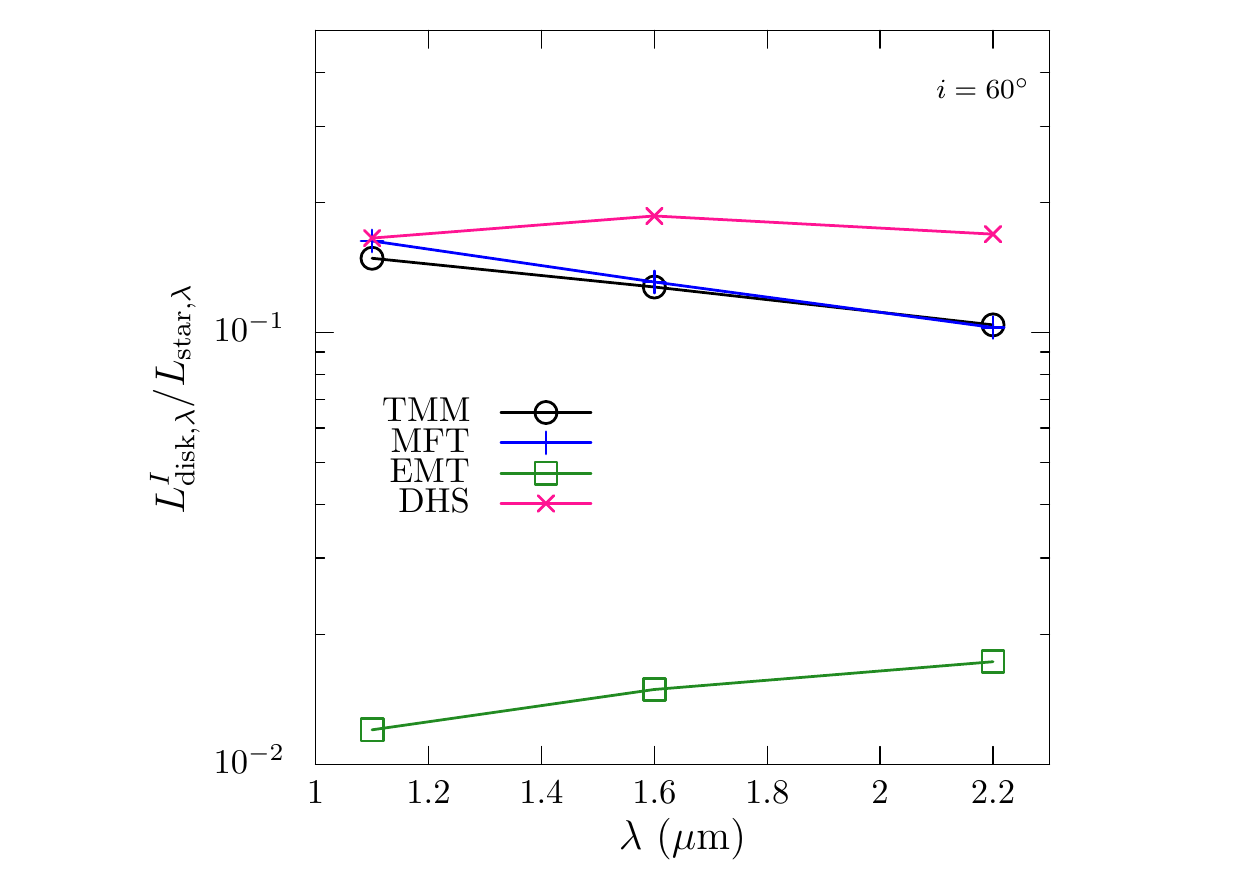}
\includegraphics[height=6.0cm,keepaspectratio]{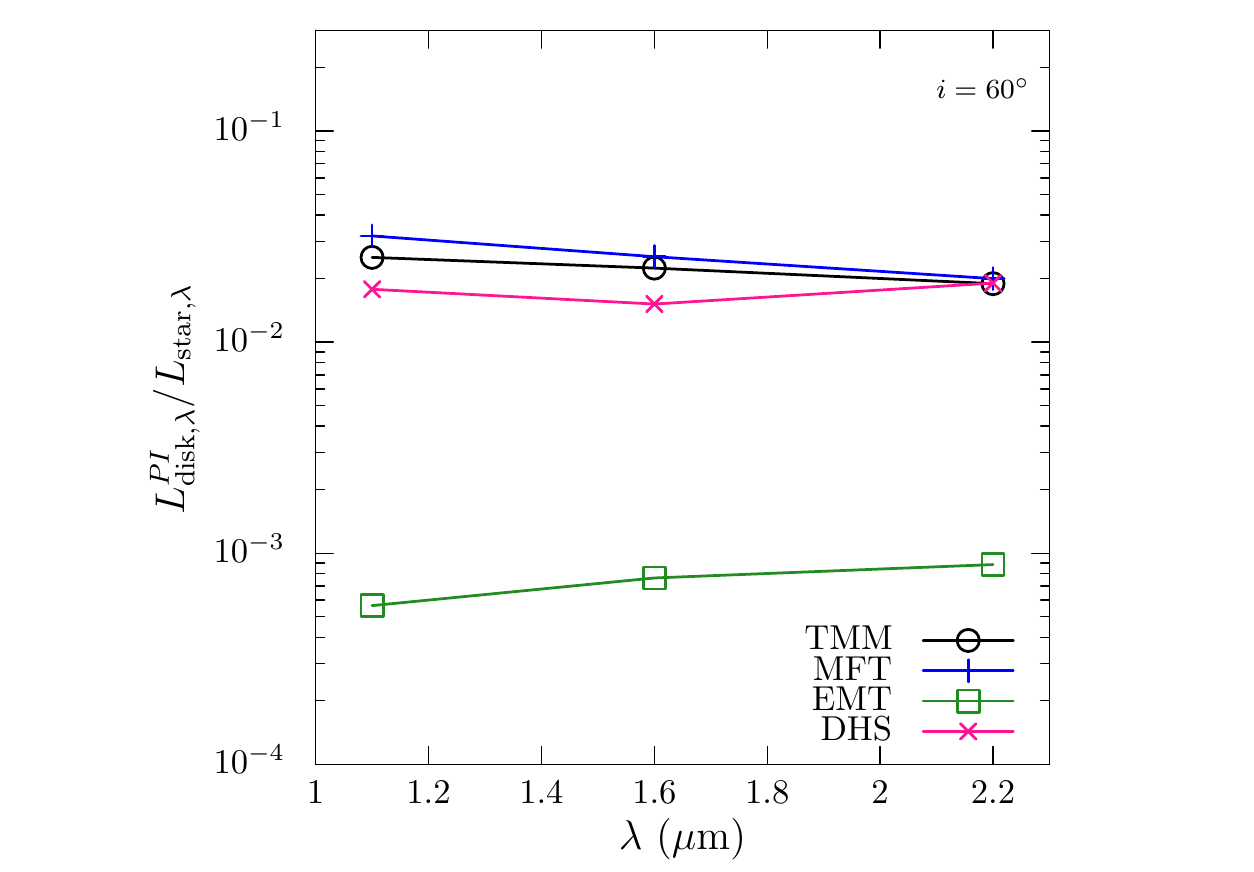}
\caption{Fractional luminosity of an inclined disc ($i=60^\circ$) for total intensity (left) and polarised intensity (right).}
\label{fig:colourincl0}
\end{center}
\end{figure*}

\begin{figure*}
\begin{center}
\includegraphics[height=13.0cm,keepaspectratio]{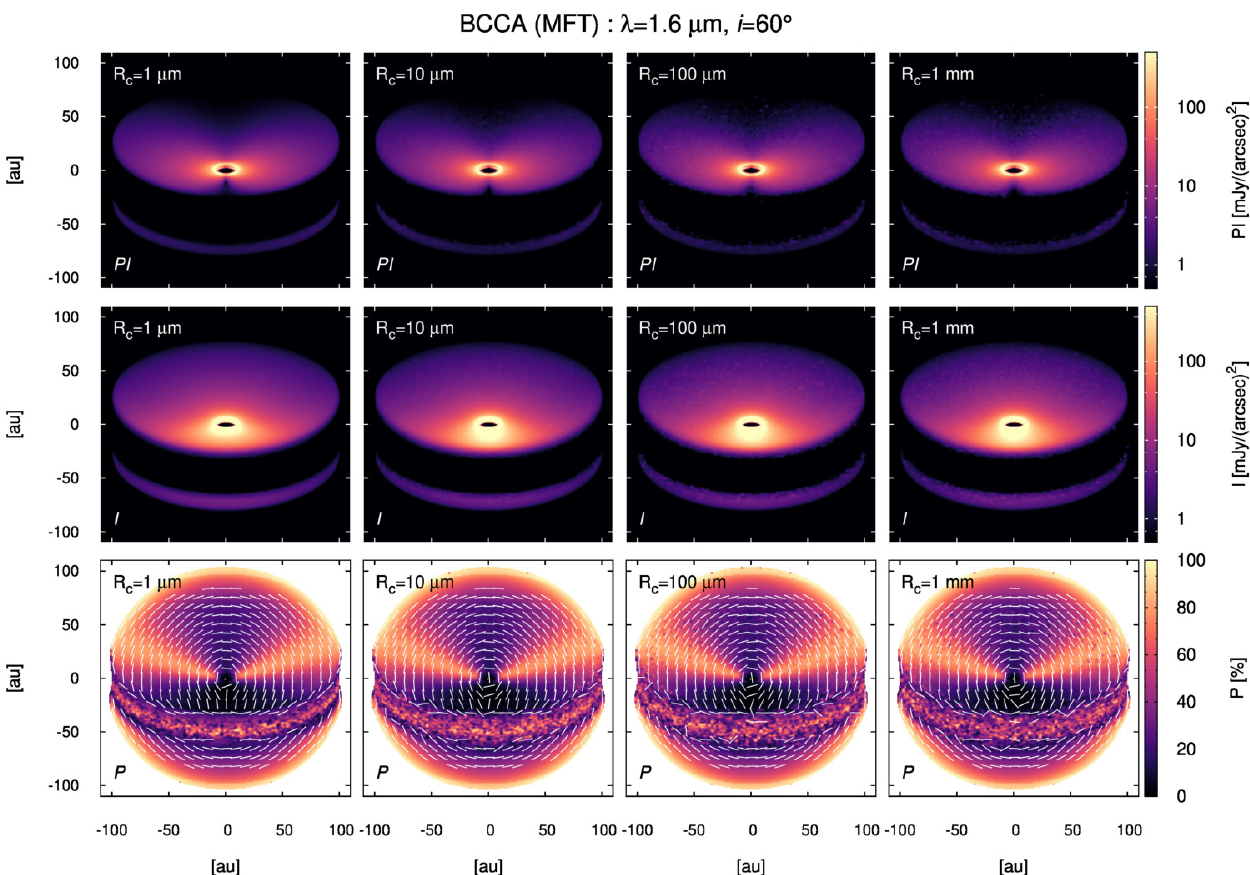}
\caption{Polarised intensity (top), total intensity (middle), and polarisation fraction (bottom). From left to right, scattered-light images of BCCA with $R_c=1\ \mu$m, $10\ \mu$m, $100\ \mu$m, and 1 mm are shown, respectively, where optical properties of BCCA are computed by using MFT.}
\label{fig:extlargeimage}
\end{center}
\end{figure*}

Figure \ref{fig:colourincl0} show scattered-light colour in total intensity and in polarised intensity calculated at inclination angle $i=60^\circ$ obtained by various approximate methods. 
Table \ref{table:colour_method} shows colours of the BCCA model for various solution methods for their optical properties.
The MFT model is able to reproduce the TMM results compared to other approximate methods. DHS shows red colour in total intensity. It is also found that once the EMT is used to obtain optical properties of large BCCA, the disc becomes faint and reddish; however, it should be strengthen that this is due to the artefact arising from the use of the Mie theory. 

Finally, we show some obtained radiative transfer images for large BCCA studied in Section \ref{sec:exlarge}. By applying MFT to BCCA with $R_c=1\ \mu$m to 1 mm, we obtain Figure \ref{fig:extlargeimage}. Because of saturation of scattering properties of BCCA (Figures \ref{fig:Z11large} and \ref{fig:saturation}), scattered light images of BCCA with various radius look similar (see also Section \ref{sec:exlarge}).

\section{Compact dust aggregates} \label{sec:rtcomp}
Scattered-light images and colours of the compact dust aggregate models with various radii are presented.

\subsection{H-band Images}
\begin{figure*}
\begin{center}
\includegraphics[height=13.0cm,keepaspectratio]{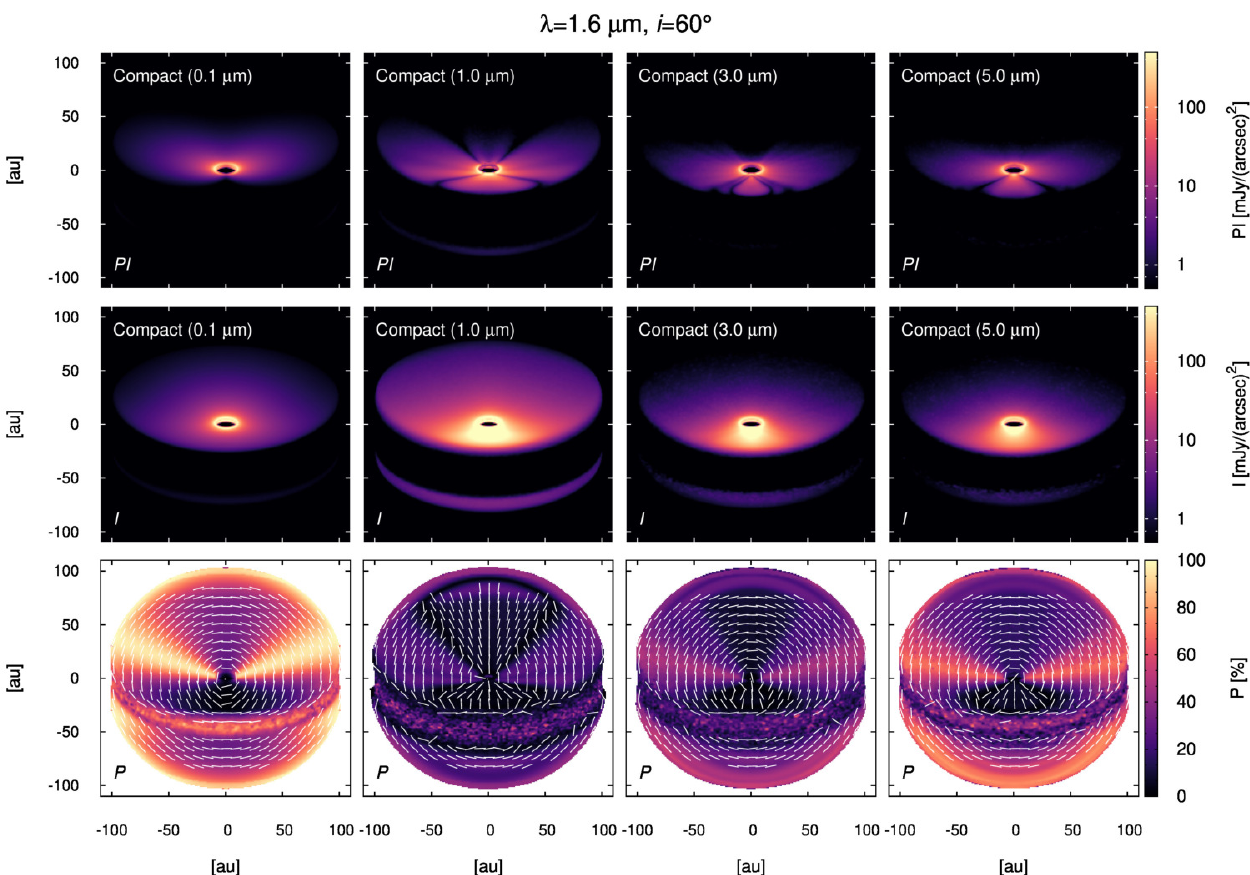}
\caption{Polarised intensity (top), total intensity (middle), and polarisation fraction (bottom). Scattered-light images of the single monomer model (leftmost) and the compact dust aggregate models (right three columns).}
\label{fig:compimages}
\end{center}
\end{figure*}

Figure \ref{fig:compimages} shows scattered-light images at $\lambda=1.6\ \mu$m for the compact dust aggregate models and the single monomer model for comparison. By using the scattered-light mapping, Figure \ref{fig:polfrac_comp} shows scattered-light intensity and polarisation fraction as a function of scattering angle at $R=$50 au.

First of all, we discuss total intensity.
When the aggregate radius is smaller than $\lambda/2\pi$, the disc is faint and shows weak front-back brightness asymmetry. Once the aggregate radius exceeds $\lambda/2\pi$, the disc begins to show front-back brightness asymmetry in total intensity. In addition, when the aggregate radius exceeds 1.0 $\mu$m, the disc total scattered light becomes faint with increasing the aggregate radius. This is because scattering phase function of large compact dust aggregate is sharply peaked at forward scattering angles, and then this results in reducing effective single scattering albedo \citep{Dullemond03, Min10, Mulders13}.

Second, we argue polarisation fraction.
When the radius of compact dust aggregate is 0.1 $\mu$m, polarisation fraction becomes more than 90\% because Rayleigh scattering happens. As the aggregate radius increases, polarisation fraction decreases. However, once the aggregate radius exceeds 1.0 $\mu$m, polarisation fraction increases again as the aggregate radius increases. This is due to the effect of Brewster scattering, which finally brings the degree of polarisation to 100\% at the Brewster angle \citep{Min16} in the limit of sufficiently large grain radius. Based on light scattering simulations of compact dust aggregates, \citet{Min16} showed that compact dust aggregates do not show high degree of polarisation at the Brewster scattering, unlike the prediction of the DHS method. The perfect polarisation at the Brewster angle is though to be arising from the ignorance of the surface roughness of compact dust aggregates in the DHS method \citep{Min16}, and hence we expect that the realistic compact dust aggregates may not show the increase of polarisation fraction at the Brewster angle.

\begin{figure*}
\begin{center}
\includegraphics[height=6.0cm,keepaspectratio]{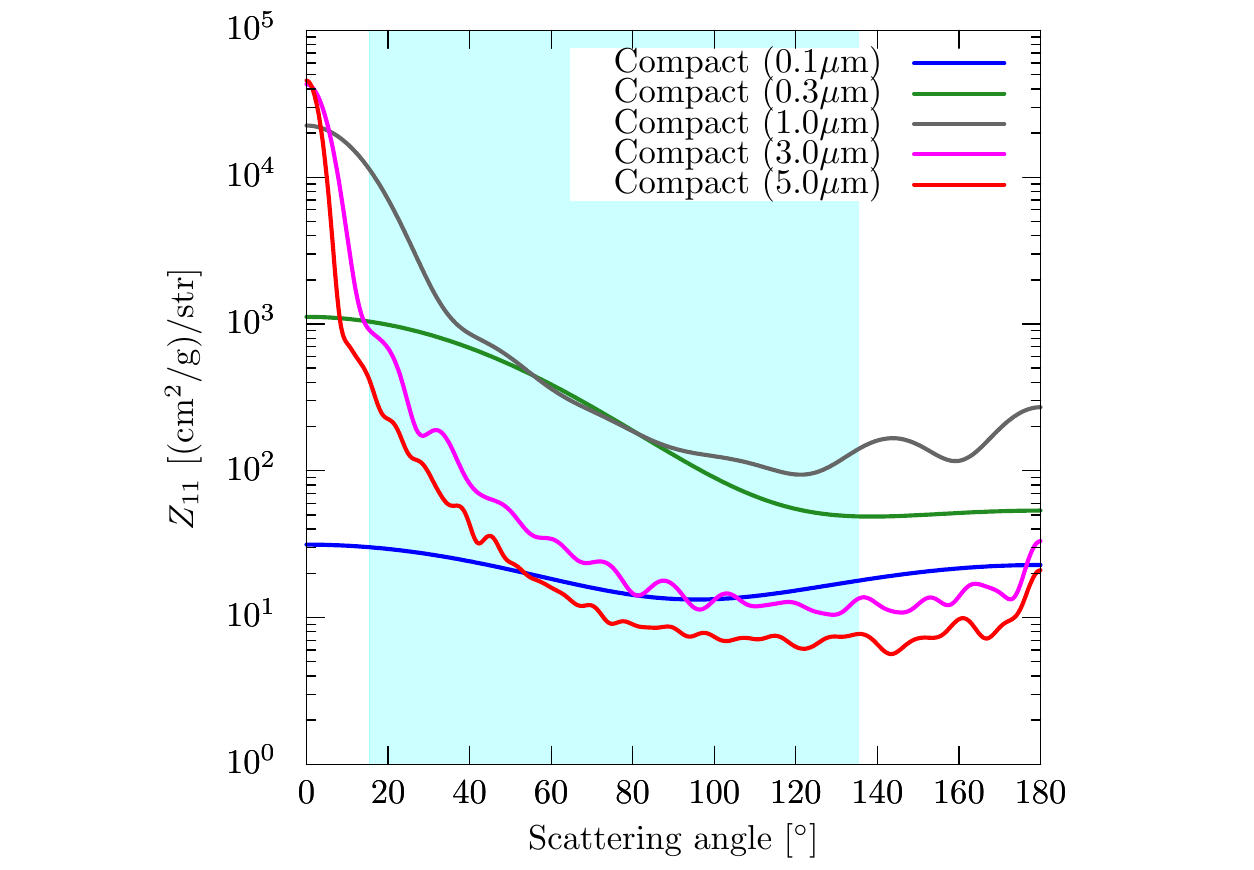}
\includegraphics[height=6.0cm,keepaspectratio]{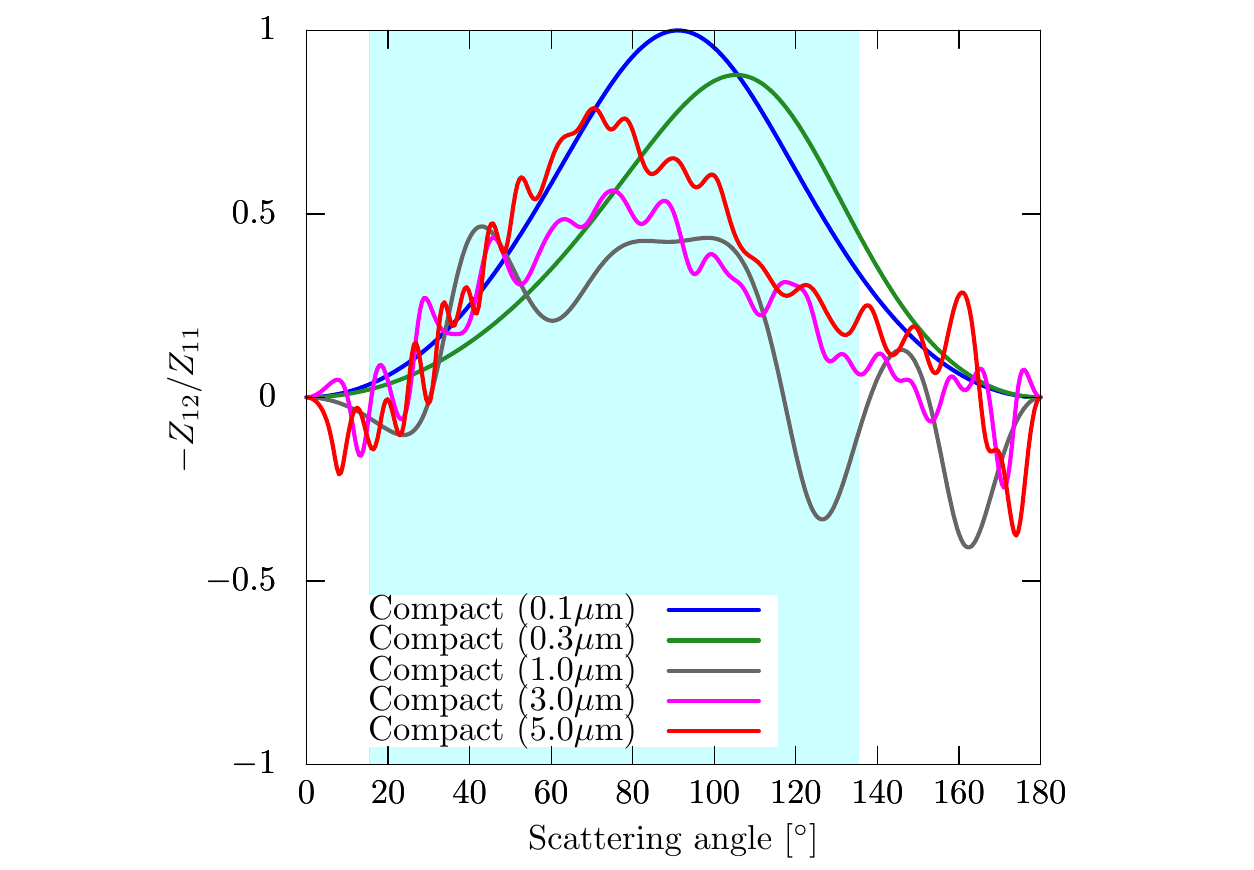}
\caption{Phase function $Z_{11}$ (left), and the degree of linear polarisation $-Z_{12}/Z_{11}$ (right). Optical properties of compact dust aggregates are obtained by DHS. Hatched region indicates a range of scattering angle to be observed for a disc with the flaring index $\beta=1.25$ and the inclination angle $i=60^\circ$.}
\label{fig:compring}
\end{center}
\end{figure*}

\begin{figure}
\begin{center}
\includegraphics[height=6.0cm,keepaspectratio]{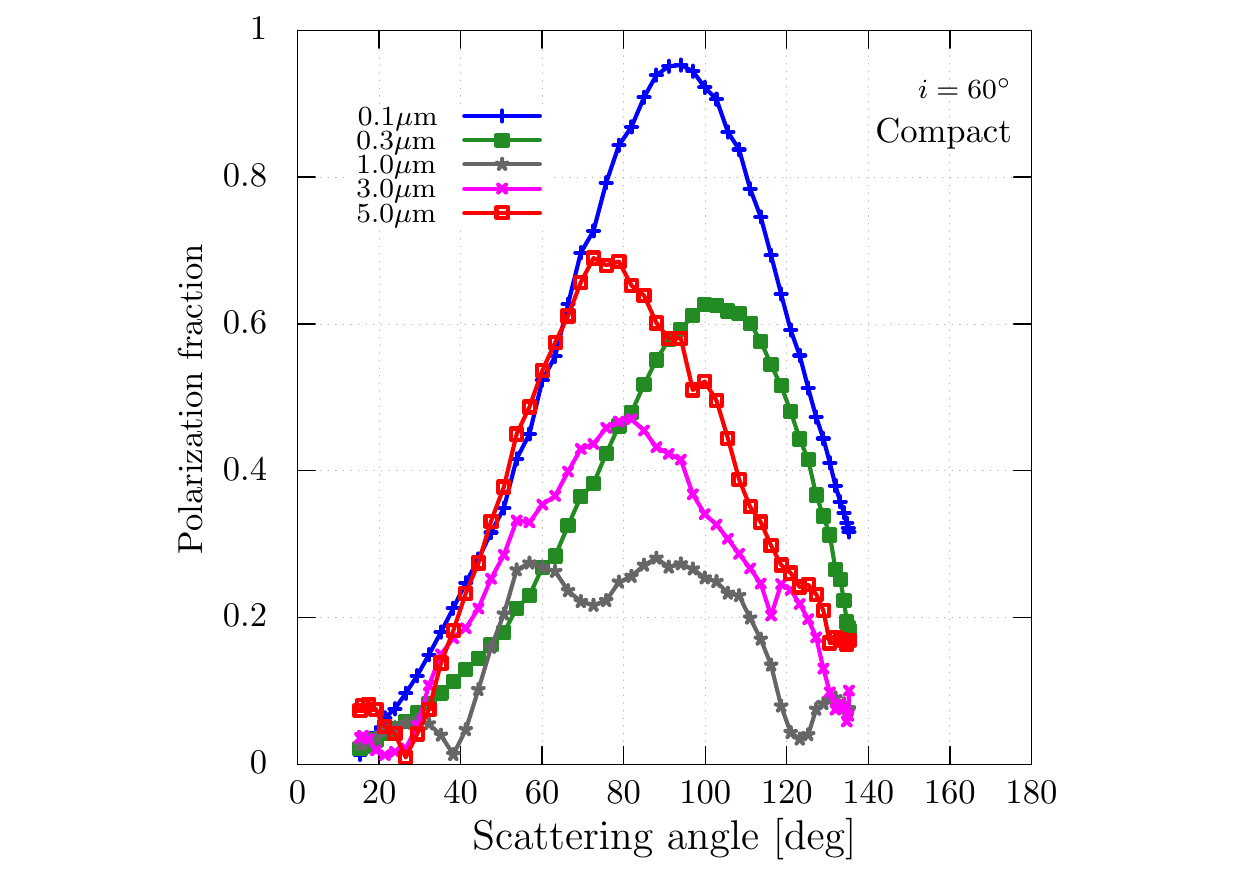}
\caption{
Polarisation fraction as a function of scattering angles at $R=50$ au measured from simulated images in Figure \ref{fig:compimages}.}
\label{fig:polfrac_comp}
\end{center}
\end{figure}

\subsection{Scattered-light colours}
\begin{figure*}
\begin{center}
\includegraphics[height=6.0cm,keepaspectratio]{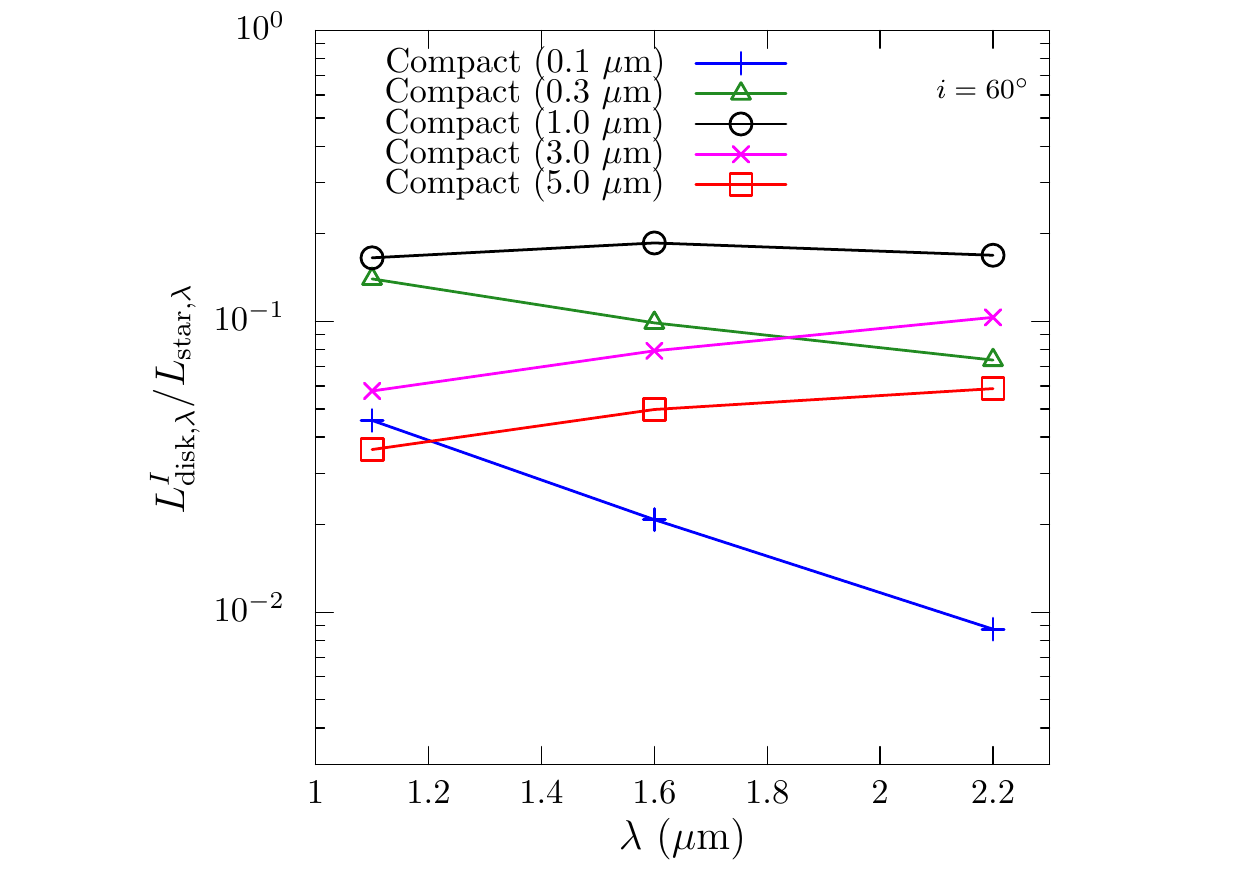}
\includegraphics[height=6.0cm,keepaspectratio]{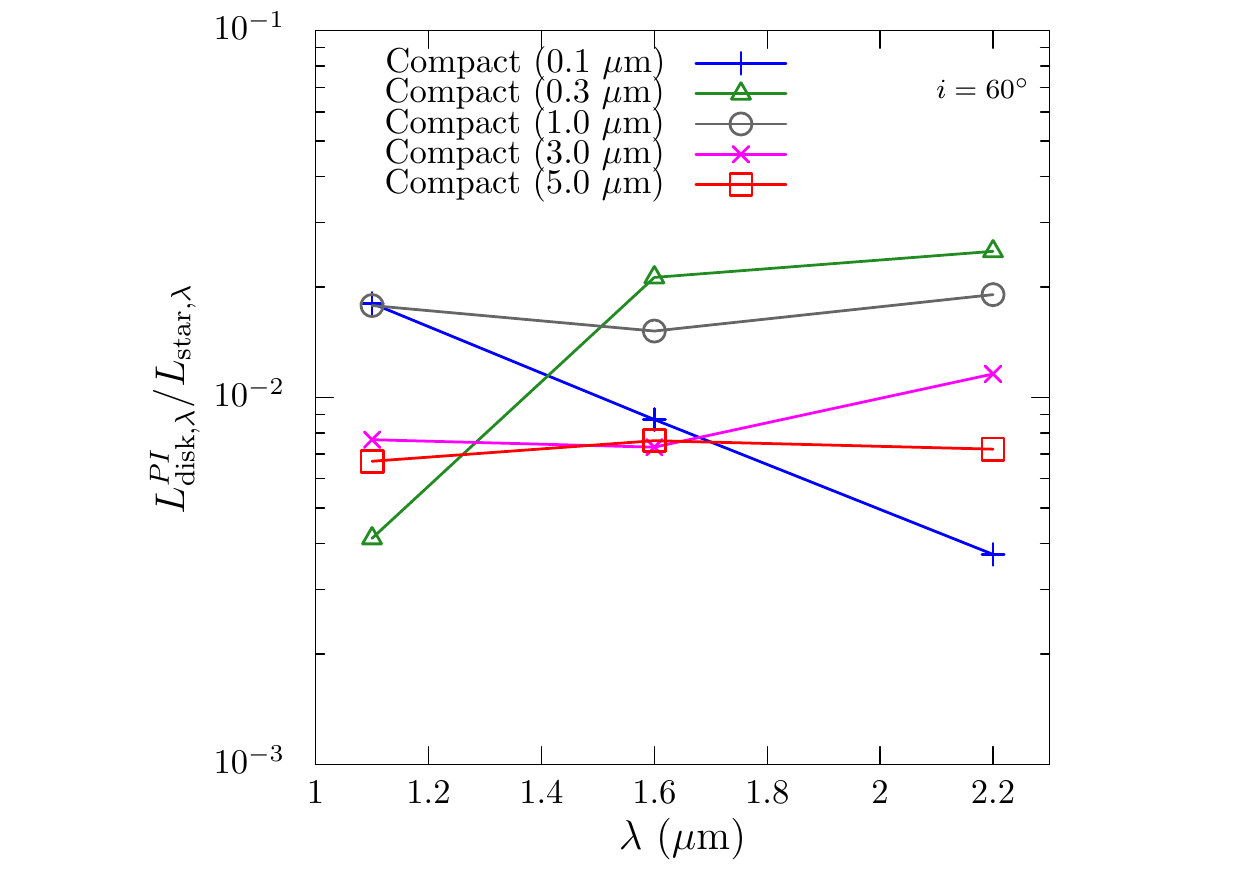}
\caption{Scattered-light colours in total intensity (left) and in polarised intensity (right) of the compact dust aggregate models. The disc inclination is assumed to be  $i=60^\circ$. Once the size parameter of the grain exceeds unity, the disc scattered light becomes faint and red as the grain radius increases.}
\label{fig:DHScolour}
\end{center}
\end{figure*}

Figure \ref{fig:DHScolour} shows disc scattered-light colour in total intensity and in polarised intensity for the compact dust aggregate models. Table \ref{table:colour_compact} shows colours of compact aggregates for various aggregate radii.

\begin{table}
  \caption{Disc Scattered Light Colour for Compact Aggregates with Various Radii}
  \label{table:colour_compact}
  \centering
  \begin{tabular}{lcc}
    \hline
    Dust model  & $\eta_\mathrm{I}$ & $\eta_\mathrm{PI}$ \\
    \hline \hline
    $0.1\ \mu$m  & $-2.4$ & $-2.3$ \\
    $0.3\ \mu$m  & $-0.93$ & $2.6$ \\
    $1.0\ \mu$m  & $0.030$ & $0.099$ \\
    $3.0\ \mu$m  & $0.84$ & $0.60$ \\
    $5.0\ \mu$m  & $0.70$ & $0.11$ \\
    \hline
 \end{tabular}
\end{table}

When the radius of the dust aggregate is smaller than the wavelength, the disc is faint and very blue in total intensity. As the radius increases, disc luminosity increases and disc colour becomes grey when the aggregate radius is comparable to the wavelength. Once the aggregate radius exceeds the wavelength, the disc becomes reddish and faint as expected by \citet{Mulders13}. 

Next, we discuss polarised intensity.
Similar to the case of total intensity, small dust aggregates show blue colours in polarised intensity. When the aggregate radius is comparable to the wavelength, polarised intensity is reddish. For the large aggregate radius, polarised intensity show grey colours; however, it is partially due to Brewster scattering. Hence, if we consider the effect of surface roughness of compact dust aggregates, polarised intensity colours might be more reddish, although more detail computations are necessary.

\section{The asymmetry parameter of sufficiently large BCCA} \label{sec:AppendixC}
The asymmetry parameter can be expressed as
\begin{equation}
g=\frac{2\pi}{k^2C_{\mathrm{sca,agg}}}\int_{-1}^1 \mu S_{11,\mathrm{agg}}(\mu)d\mu, \label{eq:def}
\end{equation}
where $C_{\mathrm{sca,agg}}$ is the scattering cross section of the dust aggregate, $k$ is the wave number, $S_{11,\mathrm{agg}}$ is a (1,1) element of scattering matrix of dust aggregates, and $\mu=\cos\theta$; where $\theta$ is the scattering angle. Using the single scattering assumption, a scattering matrix element of dust aggregates can be written by 
\begin{equation}
S_{11,\mathrm{agg}}(\mu)=N^2S_{11,\mathrm{mono}}(\mu)\mathcal{S}(q),
\end{equation}
where the magnitude of scattering vector $q=2k\sin(\theta/2)$ and $\mathcal{S}(q)$ is the structure factor \citep{Tazaki16}. When $qR_g\gg1$ and $d_f=2$, we can approximately decompose the scattering phase function of fluffy dust aggregates by the sum of coherent and incoherent contribution:
\begin{eqnarray}
S_{11,\mathrm{agg}}(\mu)&=&S_{11,\mathrm{agg}}^{\mathrm{coherent}}+S_{11,\mathrm{agg}}^{\mathrm{incoherent}}, \label{eq:coh}\\
S_{11,\mathrm{agg}}^{\mathrm{coherent}}&\simeq&N^2S_{11,\mathrm{mono}}(\mu)\delta(\mu-1),\\
S_{11,\mathrm{agg}}^{\mathrm{incoherent}}&\simeq&NS_{11,\mathrm{mono}}(\mu)[(qR_0)^{-2}+1],
\end{eqnarray}
where we have used Equation (29) of \citet{Tazaki16}.
Using Equations (\ref{eq:def} and \ref{eq:coh}) and $qR_0\ll1$, we obtain
\begin{equation}
g\simeq\frac{2\pi N}{k^2C_{\mathrm{sca,agg}}}\int_{-1}^{1} \mu(qR_0)^{-2}S_{11,\mathrm{mono}}(\mu)d\mu,\\
\end{equation}
Since scattering cross section is approximately proportional to $k^2$ \citep[][]{Berry86} and $S_{11,\mathrm{mono}}\propto k^6$ (Rayleigh scattering) \citep{Bohren83}, the asymmetry parameter $g$ becomes wavelength independent.
Therefore, sufficiently large dust aggregates ($qR_g\gg1$) with $d_f=2$ containing small monomers ($qR_0\ll1$) yield wavelength independent asymmetry parameter.


\bsp	
\label{lastpage}
\end{document}